\documentclass[numberwithinsect,cleveref,thm-restate]{no-lipics-v2019}

\usetikzlibrary{shapes.misc, decorations.pathreplacing, calc}
\usepackage{bm}

\nolinenumbers


\newcommand{\Oh}{{O}}
\newcommand{\tOh}{{\widetilde{O}}}

\newcommand{\karl}[1]{}

\newcommand{\php}[1]{}
\newcommand{\tmp}[1]{{\color{purple}}}

\title{On Near-Linear-Time Algorithms for Dense Subset Sum}

\author{Karl Bringmann}{Saarland University and Max Planck Institute for Informatics,\and
Saarland Informatics Campus, Saarbrücken, Germany}{bringmann@cs.uni-saarland.de}{}{}

\author{Philip Wellnitz}{Max Planck Institute for Informatics,\and
Saarland Informatics Campus, Saarbrücken, Germany}{wellnitz@mpi-inf.mpg.de}{https://orcid.org/0000-0002-6482-8478}{}

\authorrunning{K. Bringmann and P. Wellnitz}
\Copyright{Karl Bringmann and Philip Wellnitz}

\funding{\emph{Karl Bringmann:} This work is part of the project TIPEA that has received funding from the European Research Council (ERC) under the European Unions Horizon 2020 research and innovation programme (grant agreement No.\ 850979).}

\def\size#1{\ensuremath{|#1|}}
\def\sm#1{\ensuremath\Sigma_{#1}}
\def\ssm#1{\ensuremath\mathcal{S}_{#1}}
\def\mx#1{\textup{mx}_{#1}}
\def\mul#1{\mu_{#1}}
\def\supp#1{\ensuremath \textup{supp}_{#1}}
\def\smi#1{\ensuremath\Sigma_{#1}}

\def\mxi#1{\operatorname{mx}_{#1}}
\def\muli#1{\mu_{#1}}
\def\la#1{\lambda_{#1}}

\def\PF#1{\operatorname{PF}_{#1}}

\def\fragment#1#2{\bm{[}\,#1\,\bm{.\,.}\,#2\,\bm{]}}
\def\position#1{\bm{[}\,#1\,\bm{]}}
\makeatother

\newcommand{\kSAT}{\textsf{$k$-SAT}\xspace}
\newcommand{\SubsetSum}{\textsf{Subset Sum}\xspace}
\newcommand{\SubsetSumN}{\textsf{Subset Sum}}
\newcommand{\kSum}{\textsf{$k$-Sum}\xspace}
\newcommand{\kSUM}{\kSum}

\newcommand{\eps}{\varepsilon}

\begin{document}
\maketitle

\begin{abstract}
    In the \SubsetSum problem we are given a set of $n$ positive integers $X$ and a target
    $t$ and are asked whether some subset of $X$ sums to $t$.
    Natural parameters for this problem that have been studied in the literature are $n$
    and $t$ as well as the maximum input number $\mx{X}$ and the sum of all input numbers $\sm{X}$.
    In this paper we study the {\em dense} case of \SubsetSum, where all these parameters
    are \emph{polynomial} in $n$. In this regime, standard pseudo-polynomial algorithms
    solve \SubsetSum in polynomial time $n^{\Oh(1)}$.

    Our main question is: \emph{When can dense Subset Sum be solved in near-linear time
    $\tOh(n)$?} We provide an essentially complete dichotomy by designing improved algorithms and proving conditional lower
    bounds, thereby determining essentially all settings of the parameters $n,t,\mx{X},\sm{X}$ for
    which dense \SubsetSum is in time~$\tOh(n)$. For notational convenience we assume
    without loss of generality that $t \ge \mx{X}$ (as larger numbers can be ignored) and
    $t \le \sm{X}/2$ (using symmetry). Then our dichotomy reads as follows:
    \begin{itemize}
        \item By reviving and improving an additive-combinatorics-based approach by Galil
            and Margalit [SICOMP'91], we show that \SubsetSum is in near-linear time $\tOh(n)$ if $t
            \gg \mx{X} \sm{X}/n^2$.
        \item We prove a matching conditional lower bound: If \SubsetSum is in near-linear
            time for any setting with $t \ll \mx{X} \sm{X}/n^2$, then the Strong
            Exponential Time Hypothesis and the Strong k-Sum Hypothesis fail.
    \end{itemize}
    We also generalize our algorithm from sets to multi-sets, albeit with non-matching
    upper and lower bounds.
\end{abstract}

\clearpage

\section{Introduction}

In the \SubsetSum problem we are given a (multi-)set $X$ of $n$ positive integers and a target $t$ and want to decide whether some subset of $X$ sums to $t$.
\SubsetSum is the most fundamental NP-hard problem at the intersection of theoretical computer science, mathematical optimization, and operations research. 
It draws much of its motivation from being a hard special case of many other problems, e.g., Knapsack and Integer Programming. 
An additional modern motivation is lattice-based crypto, which is partly based upon average-case hardness of a variant of \SubsetSum (specifically the Short Integer Solution problem~\cite{Ajtai96}).

Algorithms for \SubsetSum have been studied for many decades (see, e.g., the monograph~\cite{KellererPP04}), and the literature on this problem is still flourishing (see, e.g.,~\cite{AxiotisBJTW19,BansalGNV17,Bringmann17,JW19,
KoiliarisX19}).
Maybe the most well-known algorithm for \SubsetSum is Bellman's pseudopolynomial $\Oh(nt)$-time algorithm~\cite{bellman1957dynamic}. 
This was recently improved to randomized time\footnote{We write $\tOh(T)$ for any function that is bounded by $\Oh(T \log^c T)$ for some $c>0$.} $\tOh(n+t)$~\cite{Bringmann17}; for further logfactor improvements see~\cite{JW19}. 
Using the modern toolset of fine-grained complexity theory, this improved running time was shown to be near-optimal, specifically any $t^{1-\eps} 2^{o(n)}$-time algorithm would violate the Strong Exponential Time Hypothesis~\cite{AbboudBHS19}, and a similar lower bound holds under the Set Cover Hypothesis~\cite{CyganDLMNOPSW16}. 
This essentially settles the time complexity with respect to parameters $n,t$.
Alternative parameters for \SubsetSum
are the maximum input number, which we denote by $\mx{X}$, and the sum of all input numbers, which we denote by $\sm{X}$. 
Studying \SubsetSum with respect to these parameters has been a significant effort in theoretical computer science and optimization, as illustrated by Table~\ref{tab:literature}. 
In particular, it is known that 
\SubsetSum can be solved in time $\Oh(n \cdot \mx{X})$~\cite{Pisinger99} or $\tOh(\sm{X})$~\cite{KoiliarisX19}.
The most crucial open problem in this line of research is whether \SubsetSum can be solved in time $\tOh(n+\mx{X})$, see~\cite{AxiotisBJTW19}.

This open problem illustrates that we are far from a complete understanding of the complexity of \SubsetSum with respect to the combined parameters $n,t,\mx{X},\sm{X}$.
A line of work from around 1990 \cite{chaimovich1999new,ChaimovichFG89,freiman1988extremal,
gm91,GalilM91} suggests that this complexity is fairly complicated, as it lead to the following result.
\begin{theorem}[Galil and Margalit~\cite{GalilM91}] \label{thm:GM}
  Given a set $X$ of $n$ positive integers and a target $t \le \sm{X}/2$, if\footnote{We use the notation ``$f \gg g$'' only in the informal overview of our results. We mostly use it to hide polylogarithmic factors, sometimes also to hide subpolynomial factors. Formally, for functions $f,g$ and a property $P$, we write ``if $f \gg g$ then $P$'' if the following statement is true: For any $\eps > 0$ there exists $C>0$ such that $f \ge C \cdot n^\eps \cdot g$ implies $P$.} $t \gg \mx{X} \sm{X} / n^2$ then \SubsetSum can be solved in time $\tOh(n + \mx{X}^2 / n^2)$.\lipicsEnd
\end{theorem}
For understanding the complexity of \SubsetSum with respect to the parameters $n,t,\mx{X},\sm{X}$, Galil and Margalit's algorithm provides a highly non-trivial upper bound in a complicated regime. In their paper they argue that their approach must fail outside their feasible regime. Nevertheless one can wonder: \emph{Is the optimal time complexity of Subset Sum really so complicated, or is this an artefact of Galil and Margalit's approach?} In particular, can we show a lower bound that establishes their regime to be natural? 
Note that Galil and Margalit discovered a non-trivial regime in which \SubsetSum can be solved in near-linear time $\tOh(n)$, namely when $\sm{X}/2 \ge t \gg \mx{X} \sm{X} / n^2$ and $\mx{X} = \tOh(n^{3/2})$. One can wonder: Can this near-linear time regime be extended? \emph{What is the largest possible regime in which Subset Sum is in near-linear time?}
In this paper, we provide answers to all of these questions.

\medskip
First, let us discuss the details of Galil and Margalit's result.
Note that the assumption $t \le \sm{X}/2$ is without loss of generality: For $t > \sm{X}$ the problem is trivial, and for $\sm{X}/2 < t \le \sm{X}$ any subset $Y \subseteq X$ sums to $t$ if and only if $X \setminus Y$ sums to $\sm{X}-t$, so the inputs $(X,t)$ and $(X,\sm{X}-t)$ are equivalent.
In particular, an alternative formulation is that this algorithm solves \SubsetSum very efficiently when the target lies in a feasible interval centered around $\sm{X}/2$.
Note that the feasible interval is only non-empty if $\mx{X} \ll n^2$, so this result only applies to the \emph{dense} setting of \SubsetSum, where all parameters $t,\mx{X},\sm{X}$ are \emph{polynomial} in $n$. 
Moreover, while all previously mentioned algorithms also work when $X$ is a multi-set, Galil and Margalit's algorithm really requires $X$ to be a set. 
Under these strict conditions, their algorithm gives a highly non-trivial result, that uses structural insights from additive combinatorics about arithmetic progressions in the set of all subset sums.
Since it discovers a regime where \SubsetSum can be solved surprisingly fast, it has for instance recently found use in an approximation algorithm for the Partition problem~\cite{MuchaW019}.

\renewcommand{\arraystretch}{1.2}
\begin{table}
\begin{center}
\begin{tabular}{lll}
\textbf{Reference} & \textbf{Running Time} & \textbf{Comments}\\
\hline
Bellman~\cite{bellman1957dynamic} & $\Oh(nt)$ & \\
Pisinger~\cite{Pisinger03} & $\Oh(n t / w)$ & RAM model with cells of $w$ bits \\
Pisinger~\cite{Pisinger99} & $\Oh(n\, \mx{X})$ & \\
Klinz and Woeginger~\cite{KlinzW99} & $\Oh(\sm{X}^{3/2})$ & \\
Eppstein~\cite{Eppstein97}, Serang~\cite{serang2014probabilistic} & $\tOh(n\, \mx{X})$ & data structure \\
Lokshtanov and Nederlof~\cite{LokshtanovN10} & $\Oh(n^3 t)$ & polynomial space, see also~\cite{Bringmann17} \\
Koiliaris and Xu~\cite{KoiliarisX19} & $\tOh(\sqrt{n} t + n)$ & \\
Koiliaris and Xu~\cite{KoiliarisX19} & $\tOh(t^{5/4} + n)$ & \\
Koiliaris and Xu~\cite{KoiliarisX19} & $\tOh(\sm{X})$ & \\
Bringmann~\cite{Bringmann17} & $\tOh(t + n)$ & randomized \\
Jin and Wu~\cite{JW19} & $\tOh(t + n)$ & randomized, improved logfactors \\
\end{tabular}
\end{center}
\caption{Short survey of pseudopolynomial-time algorithms for \SubsetSum on multi-sets. The input consists of a multi-set $X$ of positive integers and a target number $t$. We write $n$ for the size of $X$, $\mx{X}$ for the maximum number in $X$, and $\sm{X}$ for the sum of all numbers in $X$. 
This table does not contain the line of work~\cite{chaimovich1999new,ChaimovichFG89,freiman1988extremal,
gm91,GalilM91} leading to Galil and Margalit's algorithm (Theorem~\ref{thm:GM}), because these algorithms only work on sets.
}
\label{tab:literature}
\end{table}

We remark that the conference version of Galil and Margalit's paper~\cite{GalilM91} claims the result as stated in Theorem~\ref{thm:GM}, but does not contain all proof details. The journal version of their paper~\cite{gm91} only proves a weaker result, assuming the stricter condition $t \gg \mx{X}^{1/2} \sm{X} / n$. Nevertheless, their conference version was recently cited and used in~\cite{MuchaW019}.
It would therefore be desirable to have an \emph{accessible full proof} of Theorem~\ref{thm:GM}. In any case, we will compare the results of this paper with Theorem~\ref{thm:GM}.

\subsection{Our Contribution}

In this paper, we study \SubsetSum in the dense regime, where the parameters $t,\mx{X},\sm{X}$ are all bounded by a polynomial in $n$. In this regime, any pseudopolynomial-time algorithm solves \SubsetSum in polynomial time $\textup{poly}(n)$. Our main result is an essentially complete dichotomy that determines all settings of the parameters $n,t,\mx{X},\sm{X}$ where \SubsetSum can be solved in near-linear time $\tOh(n)$.

We start by discussing the case where $X$ is a set (not a multi-set).

\paragraph*{Algorithm}

Galil and Margalit discovered a non-trivial regime where \SubsetSum can be solved in near-linear time~$\tOh(n)$, namely when $\sm{X}/2 \ge t \gg \mx{X} \sm{X} / n^2$ and $\mx{X} = \tOh(n^{3/2})$. We extend the near-linear-time regime, specifically we remove the restriction $\mx{X} = \tOh(n^{3/2})$ from their regime.
We achieve this by following the same high-level approach as Galil and Margalit, but exchanging almost all parts of the algorithm in order to improve the running time.
Moreover, we provide a full proof which we think is easily accessible. 

\begin{theorem} \label{thm:intro_main_algo}
  Given a set $X$ of $n$ positive integers and a target $t \le \sm{X}/2$, if $t \gg \mx{X} \sm{X} / n^2$ then \SubsetSum can be solved in time $\tOh(n)$.\lipicsEnd
\end{theorem}

Our $\gg$-notation hides the same number of logfactors and comparable constants in Theorems~\ref{thm:GM} and~\ref{thm:intro_main_algo}.
We also remark that the recent trend of additive-combinatorics-based algorithm design typically leads to improved, but nasty running times~\cite{ChanL15,BringmannN20,MuchaW019}, so our clean running time of $\tOh(n)$ is an exception.

\paragraph*{Conditional Lower Bound}

We prove a lower bound based on the standard Strong Exponential Time Hypothesis~\cite{ip01,cip09} from fine-grained complexity theory. Alternatively, our bound can be based on the (less standard) Strong k-Sum Hypothesis~\cite{AmirCLL14,AbboudBBK17}. For details on these hypotheses, see Section~\ref{sec:lowerbound}.

\begin{theorem}[Informal] \label{thm:intro_lowerbound}
  \SubsetSum requires time $(\mx{X}\sm{X}/(nt))^{1-o(1)}$, unless the Strong Exponential Time Hypothesis and the Strong k-Sum Hypothesis both fail. This even holds when $X$ must be a set.\lipicsEnd
\end{theorem}

More precisely, we prove this lower bound for any {\em parameter setting} of $n,t,\mx{X},\sm{X}$ (similar to~\cite{BringmannK18}). Specifically, for the parameters $t,\mx{X},\sm{X}$ we fix corresponding exponents $\tau,\xi,\sigma\in \mathbb{R}$ and focus on instances with $t = \Theta(n^\tau)$, $\mx{X} = \Theta(n^\xi)$, and $\sm{X} = \Theta(n^\sigma)$. Some settings of $\tau,\xi,\sigma$ are trivial, in the sense that they admit no (or only finitely many) instances; we ignore such settings. For each non-trivial parameter setting, we prove a conditional lower bound of $(\mx{X}\sm{X}/(nt))^{1-o(1)}$. This shows that we did not miss any setting in which \SubsetSum admits algorithms running in time $(\mx{X}\sm{X}/(nt))^{1-\Omega(1)}$.

\smallskip
Note that for $t \ll \mx{X} \sm{X} / n^2$ we obtain a super-linear lower bound. This complements our Theorem~\ref{thm:intro_main_algo}, which runs in near-linear time $\tOh(n)$ if $t \gg \mx{X} \sm{X} / n^2$.
In particular, we obtain an \emph{essentially complete dichotomy} of near-linear-time settings, except for leaving open settings with $t \approx \mx{X} \sm{X} / n^2$. 
That is, we determined \emph{the largest possible regime in which dense Subset Sum is in near-linear time}.

Also note that our lower bound establishes the regime $t \gg \mx{X} \sm{X} / n^2$ to be \emph{natural}, and not just an artefact of the algorithmic approach, so the optimal time complexity of \SubsetSum is indeed complicated!

\paragraph*{Multi-Sets}

Finally, we provide a generalization of our algorithm (and thus also Galil and Margalit's result) to multi-sets.
For a multi-set $X$, we denote by $\mul{X}$ the largest multiplicity of any number in $X$. 

\begin{theorem} \label{thm:intro_main_multi}
  Given a multi-set $X$ of $n$ positive integers and a target $t \le \sm{X}/2$, if $t \gg \mul{X} \mx{X} \sm{X} / n^2$ then \SubsetSum can be solved in time $\tOh(n)$.\lipicsEnd
\end{theorem}

For constant multiplicity $\mul{X} = \Oh(1)$ this yields the same result as for sets. For larger multiplicities, we pay a factor $\mul{X}$ in the feasibility bound.
Since any set is also a multi-set, the lower bound from Theorem~\ref{thm:intro_lowerbound} also applies here. However, for $\mul{X} \gg 1$ we no longer obtain matching regimes.

\subsection{Organization}

After formalizing some notation in Section~\ref{sc:notation}, we give a technical overview of our results in Section~\ref{sec:techoverview}.
We present our algorithm in Section~\ref{sec:algo}, and our conditional lower bound in Section~\ref{sec:lowerbound}.
Finally, we conclude with open problems in Section~\ref{sec:conclusion}.

\section{Notation}\label{sc:notation}

We write $\position{n} := \{1,\dots,n\}$ and $\fragment{\ell}{r} := \{\ell,\dots,r\}$. 
Further, for a set $A$ and an integer $d$,
we write $A \bmod d := \{a \bmod d \mid a \in A\}$.

Throughout the paper we let $X$ denote a \emph{finite non-empty multi-set of positive integers}
(or \emph{multi-set} for short). For an integer $x$, we write
$\mu(x;X)$ to denote the multiplicity of $x$ in $X$. A number that does not appear in~$X$ has
multiplicity 0. We use the same notation for multi-sets and sets, e.g., a subset $Y \subseteq X$ is a multi-set with $\mu(x;Y) \le \mu(x;X)$ for all $x$.
We write $\supp{X}$ to denote the support of $X$, that is, the set of all distinct
integers contained in the multi-set~$X$.

We associate the following relevant objects to a multi-set $X$:
\begin{itemize}
    \item \emph{Size} $|X|$: The number of elements of $X$, counted with multiplicity,
        that is, $|X| := \sum_{x \in \mathbb{N}} \mu(x;X)$.
    \item \emph{Maximum} $\mx{X}$: The maximum element of $X$, that is, $\mx{X} := \max\{x \in
        \mathbb{N} \mid \mu(x;X) > 0\}$.
    \item \emph{Multiplicity} $\mul{X}$: The maximum multiplicity of $X$, that is,
        $\muli{X} := \max\{\mu(x;X) \mid x \in \mathbb{N}\}$.
    \item \emph{Sum} $\sm{X}$: The sum of all elements of $X$, that is, $\sm{X} := \sum_{x \in
        \mathbb{N}} x \cdot \mu(x;X)$.
    \item \emph{Set of all subset sums} $\ssm{X}$: The set containing all sums of subsets
        of $X$, that is, $\ssm{X} := \{\sm{Y} \mid Y \subseteq X\}$.
\end{itemize}
The \SubsetSum problem now reads as follows.
\begin{problem}[\sf Subset Sum]
    Given a (multi-)set $X$ and an integer $t$, decide whether $t \in \ssm{X}$.\lipicsEnd
\end{problem}

For a multi-set $X$ and an integer $k \ge 1$ we write $kX := \{ kx\mid x\in X\}$, that is, every number in $X$ is multiplied by $k$.
Similarly, if every number in $X$ is divisible by $k$, then we write $X/k := \{ x/k\mid x\in X\}$.

\section{Technical Overview} \label{sec:techoverview}

\subsection{Technical Overview of the Algorithm}

We follow the same high-level approach as Galil and Margalit~\cite{GalilM91}. However, we replace essentially every part of their algorithm to obtain our improved running time as well as a generalization to multi-sets. 

Our goal is to design a near-linear time algorithm for \SubsetSum in the regime $t \gg \mul{X} \mx{X} \sm{X} / |X|^2$. Recall that without loss of generality we can assume $t \le \sm{X}/2$, by using symmetry.
Combining these inequalities, we will in particular assume $|X|^2 \gg \mul{X} \mx{X}$. We formalize this assumption as follows.

\begin{restatable}[Density]{definition}{defdensity} \label{df:den}
    We say that a multi-set $X$ is {\em $\delta$-dense} if it satisfies $|X|^2 \ge
    \delta\cdot\mul{X}\cdot \mx{X}$.\lipicsEnd
\end{restatable}

If (almost) all numbers in $X$ are divisible by the same integer $d>1$, then it may be that not all remainders modulo $d$ are attainable by subsets sums. For this reason, we introduce the following notion. 

\begin{restatable}[Almost Divisor]{definition}{defalmostdivisor} \label{df:ald}
    We write $X(d) := X \cap d \mathbb{Z}$ to denote the multi-set of all numbers in~$X$ that are
  divisible by $d$. Further, we write $\overline{X(d)} := X \setminus X(d)$ to denote the
  multi-set of all numbers in $X$ not divisible by $d$.
  We say an integer $d > 1$ is an {\em $\alpha$-almost divisor} of $X$ if $|\overline{X(d)}| \le
  \alpha \cdot \mul{X}\cdot \sm{X} / |X|^2$.\lipicsEnd
\end{restatable}

Using the above definitions, we can cleanly split our proof into a {\em structural part} and an
{\em algorithmic part}. 
We first formulate these two parts and show how they fit together to solve \SubsetSum. We then discuss the proofs of our structural and algorithmic part in Sections~\ref{sec:techoverview_struct} and \ref{sec:techoverview_algo} below.

\smallskip
In the structural part we establish that dense and almost-divisor-free sets generate all possible subset sums apart from a short prefix and suffix. Note that any $X$ satisfies $\ssm{X} \subseteq \fragment{0}{\sm{X}}$.

\begin{restatable}[Structural Part, Informal]{theorem}{thmmainstruct} \label{thm:main_struct}
  If $X$ is $\widetilde{\Theta}(1)$-dense\footnote{In this technical overview we present informal versions of our intermediate theorems. In particular, we write $\widetilde{\Theta}(1)$ to hide a \emph{sufficiently large} polylogarithmic factor $C \log^C(n)$. These factors are made precise later in the paper.} and has no $\widetilde{\Theta}(1)$-almost divisor, then there exists $\la{X} = \widetilde{\Theta}(\mul{X} \mx{X} \sm{X} / |X|^2)$ such that $\fragment{\la{X}}{\sm{X} - \la{X}} \subseteq
  \ssm{X}$.\lipicsEnd
\end{restatable}

Our algorithmic part is a reduction of the general case to the setting of Theorem~\ref{thm:main_struct}. This is achieved by repeatedly removing almost divisors (i.e., finding an almost divisor $d$ and replacing $X$ by $X(d)/d$). 

\begin{restatable}[Algorithmic Part, Informal]{theorem}{thmmainalgo} \label{thm:main_algo}
  Given an $\widetilde{\Theta}(1)$-dense multiset $X$ of size $n$, in time $\tOh(n)$ we can compute an integer $d \ge 1$ such that $X' := X(d)/d$ is $\widetilde{\Theta}(1)$-dense and has no $\widetilde{\Theta}(1)$-almost divisor. \lipicsEnd
\end{restatable}

By combining these two components, we show that in the regime $t \gg \mul{X} \mx{X} \sm{X} / |X|^2$ the \SubsetSum problem is characterized by its behaviour modulo $d$.

\begin{theorem}[Combination I, Informal] \label{thm:combine_struct}
  Let $X$ be an $\widetilde{\Theta}(1)$-dense multi-set, and let $d$ be as in Theorem~\ref{thm:main_algo}. Then for any $t \le \sm{X}/2$ with $t \gg \mul{X} \mx{X} \sm{X} / |X|^2$ we have 
  \[ t \in \ssm{X} \quad\text{if and only if}\quad t \bmod d \in \ssm{X} \bmod d. \]
\end{theorem}
\begin{proof}[Proof Sketch]
In one direction, if $t \bmod d \not\in \ssm{X} \bmod d$, then clearly $t$ is not a subset sum of $X$. In the other direction, if $t \bmod d \in \ssm{X} \bmod d$, then there is a subset $Y \subseteq X$ summing to $t$ modulo $d$. We can assume that $Y \subseteq \overline{X(d)}$, since numbers divisible by $d$ do not help for this purpose.
The remaining target $t' = t - \sm{Y}$ is divisible by $d$. Since we assume $t$ to be large and by arguing about the size of $Y$, we can show that $t'/d$ lies in the feasible interval of Theorem~\ref{thm:main_struct} applied to $X(d)/d$. Thus, some subset of $X(d)/d$ sums to $t'/d$, meaning some subset of $X(d)$ sums to $t'$. Together, we have found a subset of $X$ summing to $t$. Hence, we can decide whether $t$ is a subset sum of $X$ by deciding the same modulo $d$. 
\end{proof}

Using the above structural insight for algorithm design yields the following result.

\begin{theorem}[Combination II, Informal] \label{thm:combine_algo}
  We can preprocess a given $\widetilde{\Theta}(1)$-dense multi-set~$X$ of size $n$ in time~$\tOh(n)$. Given a query $t \le \sm{X}/2$ with $t \gg \mul{X} \mx{X} \sm{X} / n^2$ we can then decide $t \in \ssm{X}$ in time $\Oh(1)$.
  
  In particular, given a multi-set $X$ of size $n$ and a target $t \le \sm{X}/2$ with $t \gg \mul{X} \mx{X} \sm{X} / n^2$ we can decide whether $t \in \ssm{X}$ in time $\tOh(n)$.
\end{theorem}
\begin{proof}[Proof Sketch]
  The preprocessing has two steps: (1)~Computing the number $d$ from Theorem~\ref{thm:main_algo}. This can be done in time $\tOh(n)$ by Theorem~\ref{thm:main_algo}. (2) Solving \SubsetSum modulo $d$, that is, computing the set $\ssm{X} \bmod d$. Here we use a recent algorithm by Axiotis et al.~\cite{AxiotisBJTW19,AxiotisBBJNTW20} that runs in time $\tOh(n+d)$, which can be bounded by $\tOh(n)$ in our context using the density assumption. 
  
  On query $t$ it suffices to check whether $t \bmod d$ lies in the precomputed set $\ssm{X} \bmod d$, by Theorem~\ref{thm:combine_struct}.
  
  For the second formulation, we argue that the assumptions $t \le \sm{X}/2$ and $t \ge \widetilde{\Theta}(\mul{X} \mx{X} \sm{X} / n^2)$ imply that $X$ is $\widetilde{\Theta}(1)$-dense. Therefore, the first formulation implies the second.
\end{proof}

It remains to describe the two main components: the structural part and the algorithmic part.

\subsubsection{Structural Part}
\label{sec:techoverview_struct}

Recall that in the structural part we analyze the setting of dense and almost-divisor-free multisets.

{
\renewcommand{\thetheorem}{\ref{thm:main_struct}}
\begin{theorem}[Structural Part, Informal]
  If $X$ is $\widetilde{\Theta}(1)$-dense and has no $\widetilde{\Theta}(1)$-almost divisor, then there exists $\la{X} = \widetilde{\Theta}(\mul{X} \mx{X} \sm{X} / |X|^2)$ such that $\fragment{\la{X}}{\sm{X} - \la{X}} \subseteq
  \ssm{X}$.\lipicsEnd
\end{theorem}
\addtocounter{theorem}{-1}
}

Note that the density assumption implies $\la{X} = o(\sm{X})$, so the interval $\fragment{\la{X}}{\sm{X} - \la{X}}$ is non-trivial.
Also note that if all numbers in $X$ were even, then all subset sums would be even, so we cannot have $\fragment{\la{X}}{\sm{X} - \la{X}} \subseteq \ssm{X}$ --- therefore it is natural to exclude almost divisors.

Theorem~\ref{thm:main_struct} is an existential result about an arithmetic progression (of stepsize 1) in the set of all subset sums, and thus belongs to the realm of \emph{additive combinatorics}, see~\cite{tao2006additive} for an overview.
Arithmetic progressions in the set of all subsets sums have been studied at least since early work of Alon~\cite{alon1987subset}, see 
also the literature by Erd{\H{o}}s, Freiman, S{\'a}rk{\"o}zy, Szemer{\'e}di, and others~\cite{alon1988sums,erdHos1990two,freiman1993new,
lev1997optimal,
l03,lipkin1989representation,s89,s94,szemeredi2006long}. A result of this type is also implicit in the algorithm by Galil and Margalit~\cite{gm91}, but only applies to sets. The main novelty of Theorem~\ref{thm:main_struct} over previous work in additive combinatorics is that we consider multi-sets with a bound on the multiplicity $\mul{X}$, which has not been explicitly studied before.

\begin{proof}[Proof Sketch]
    The proof is an elementary, but involved construction using arguments that are standard in the additive combinatorics community, but not in theoretical computer science. 

    The multi-set $X$ is partitioned into three suitable subsets $A \cup R
    \cup G$ such that:
    \begin{itemize}
        \item $\ssm{A}$ contains a long arithmetic progression of small
            step size $s$ and small starting value. To construct~$A$, we adapt the proof of a result by S{\'a}rk{\"o}zy~\cite{s94}, to generalize it from sets to multi-sets. 
        \item $R$ generates all remainders modulo $s$, that is,
            $\ssm{R} \bmod s = \mathbb{Z}_s$. To construct $R$, it suffices to pick any $s$ numbers in $\overline{X(s)}$. (We discuss later how to avoid that all numbers in $\overline{X(s)}$ are already picked by $A$.)
        \item The remaining elements $G = X \setminus (A \cup R)$ still have a large sum.
    \end{itemize}
    Using this partitioning, for any target $t \in \fragment{\la{X}}{\sm{X}-\la{X}}$ we construct a set summing to $t$ as follows. We first greedily pick elements from $G$ that sum to a number $t' = t - \Theta(\la{X})$. We then pick elements from~$R$ that sum to $t - t'$ modulo $s$. It remains to add the right multiple of $s$. This number appears as an element of the arithmetic progression guaranteed by $A$, so we pick the corresponding subset of $A$.
\end{proof}

\begin{remark}
Our proof of Theorem~\ref{thm:main_struct} is constructive and yields a polynomial-time algorithm.
    However, we currently do not know how to obtain near-linear time preprocessing and
    solution reconstruction. Fortunately, for the decision version of Subset Sum an existential result suffices.\lipicsEnd
\end{remark}

\subsubsection{Algorithmic Part}
\label{sec:techoverview_algo}

Recall that our algorithmic part is a reduction to the almost-divisor-free setting.

\thmmainalgo*

A similar result is implicit in the algorithm by Galil and Margalit~\cite{gm91}. However, their running time is $\tOh(n + (\mx{X}/n)^2)$, which ranges from $\tOh(n)$ to $\tOh(n^2)$ in our near-linear-time regime.
The main difference is that they compute $d$ using more or less brute force, specifically the bottleneck of their running time is to test for every
    integer $1 < d \le \mx{X}/n$ and for each of the $\Oh(\mx{X}/n)$ smallest elements $x
    \in X$ whether $d$ divides $x$. In contrast, we read off almost divisors from the prime factorizations of the numbers in $X$.
Another difference is that they construct $d$ by a direct method, while we iteratively construct $d =d_1 \cdots d_i$.

\begin{proof}[Proof Sketch]
    Consider the following iterative procedure. Initialize $X_0 := X$ and $i=1$. While $X_{i-1}$ has an almost divisor, we pick any almost divisor $d_i$ of $X_{i-1}$, and we continue with $X_{i} := X_{i-1}(d_i)/d_i$. The final set $X_i = X(d_1\cdots d_i)/(d_1 \cdots d_i)$ has no almost divisor, so we return $d := d_1\cdots d_i$.

    We need to show that the resulting set $X_i$ is $\widetilde{\Theta}(1)$-dense. 
    The key step here is to establish the size bound $|X_i| = \Omega(n)$. This allows us to control all relevant parameters of~$X_i$. We thus obtain existence of a number $d$ with the claimed properties.

    It remains to show that this procedure can be implemented to run in time $\tOh(n)$. The number of iterations is $\Oh(\log n)$, since the product $d_1 \cdots d_i$ grows exponentially with $i$. Therefore, the running time is dominated by the time to find an almost divisor $d_i$, if there exists one. We observe that if there exists an almost divisor, then there exists one that is a \emph{prime} number. It would thus be helpful to know the prime factorizations of all numbers in $X$. Indeed, from these prime factorizations we could read off all primes that divide sufficiently many elements of $X$, so we could infer all prime almost divisors.
    As it turns out (see Theorem~\ref{thm:PF} below), we can simultaneously factorize all numbers in $X$ in total time $\tOh(n + \sqrt{\mx{X}})$. This can
    be bounded by $\tOh(n)$ using that
    $X$ is $\widetilde{\Theta}(1)$-dense. It follows that our procedure can be implemented in time $\tOh(n)$. 
\end{proof}

The above algorithm crucially relies on computing the prime factorization of all input numbers.

\begin{restatable}{theorem}{thmPF} \label{thm:PF}
  The prime factorization of $n$ given numbers in $\position{s}$ can be computed in time
    $\tOh(n + \sqrt{s})$.
\end{restatable}

In the proof of Theorem~\ref{thm:main_algo}, we use this algorithm for $s = \Oh(n^2)$, where it
runs in time $\tOh(n)$.
From the literature (see, e.g.,~\cite{crandall2006prime}), we know three
alternatives to our algorithm, which are all worse for us:
\begin{itemize}
    \item After constructing the Sieve of Eratosthenes on $\position{s}$ in time $\tOh(s)$, we can
        determine all prime factors of a number in $\position{s}$ in time $\Oh(\log s)$. This
        yields a total running time of $\tOh(s+n)$.
    \item The prime factorization of a number in $\position{s}$ can be computed in expected time
        $s^{o(1)}$ (more precisely, time $2^{\Oh((\log s)^{1/2} (\log \log s)^{1/2})}$ for
        rigorously analyzed algorithms~\cite{lenstra1992rigorous}, and time $2^{\Oh((\log
        s)^{1/3} (\log \log s)^{2/3})}$ for heuristics, see~\cite{pomerance2008tale}).
        Running this for each of $n$ input numbers takes expected time $n \cdot s^{o(1)}$.
        For $s = \Oh(n^2)$, we improve upon this running time by a factor $s^{o(1)}$, and
        our algorithm is \emph{deterministic}.
    \item The fastest known deterministic factorization algorithms are due to
        Pollard~\cite{pollard1974theorems} and Strassen~\cite{strassen1976einige} and
        factorize a number in $\position{s}$ in time $\tOh(s^{1/4})$. Running this for all $n$
        input numbers takes time $\tOh(n \cdot s^{1/4})$. For $s \le n^2$, we improve this
        running time by a factor $s^{1/4}$.
\end{itemize}

\begin{proof}[Proof Sketch]
    Suppose we want to factorize $m_1,\ldots,m_n \in \position{s}$.
    Let $p_1,\ldots,p_\ell$ denote all primes below~$\sqrt{s}$.
    Their product $P = p_1 \cdots p_\ell$ is an $\tOh(\sqrt{s})$-bit number.
    We compute $P$ in a bottom-up tree-like fashion; this takes time
    $\tOh(\sqrt{s})$.
    Similarly, we compute $M = m_1 \cdots m_n$ in a bottom-up tree-like fashion; this
    takes time $\tOh(n)$.
    We can now compute $P \bmod
    M$.
    Then we iterate over the same tree as for $M$ in a \emph{top-down} manner, starting from the value $P \bmod
    M$ at the root and computing the values $P \bmod m_j$ at the leaves; this again can be done in time~$\tOh(n)$.
    From these values we compute the greatest common divisor of $P$ and $m_j$ as $\textup{gcd}(P, m_j) = \textup{gcd}(P \bmod m_j, m_j)$.
    Observe that $m_j > \sqrt{s}$ is prime if and only if $\textup{gcd}(P, m_j) = 1$, so we can now filter out primes.

    For composites, we repeat the above procedure once with the left half of the
    primes $p_1,\ldots,p_{\ell/2}$ and once with the right half $p_{\ell/2+1},\ldots,p_\ell$. We can infer which composites $m_j$ have a prime factor
    among the left half, and which have a prime factor among the right half. We then recurse on these halves. In the base case we find prime factors. 
\end{proof}

\subsection{Technical Overview of the Conditional Lower Bound}

Our goal in the lower bound is to show that \SubsetSum cannot be solved in near-linear time for $t \ll \mx{X} \sm{X} / n^2$, in the case where $X$ is a set. To this end, we present a conditional lower bound in the realm of fine-grained complexity theory.

We start by defining \emph{parameter settings}: For the parameters $t,\mx{X},\sm{X}$ we fix corresponding exponents $\tau,\xi,\sigma \in \mathbb{R}_+$ and consider \SubsetSum instances $(X,t)$ satisfying $t = \Theta(n^\tau)$, $\mx{X} = \Theta(n^\xi)$, and $\sm{X} = \Theta(n^\sigma)$. We call the family of all these instances a \emph{parameter setting} and denote it by $\SubsetSum(\tau,\xi,\sigma)$. 
Note that some choices of the exponents $\tau,\xi,\sigma$ are contradictory, leading to trivial parameter settings that consist of only finitely many instances or that can otherwise be solved trivially. For example, if $X$ is a set of size $n$ then $\sm{X} \ge \sum_{i=1}^n i = \Theta(n^2)$, so parameter settings with $\sigma < 2$ are trivial.
Accordingly, we call a parameter setting $(\tau,\xi,\sigma)$ \emph{non-trivial} if it satisfies the inequalities $\sigma \ge 2$ as well as $1 \le \xi \le \tau \le \sigma \le 1+\xi$; for a justification of each one of these restrictions see Section~\ref{sec:lb:lb}.

For every non-trivial parameter setting $\SubsetSum(\tau,\xi,\sigma)$, we prove a conditional lower bound ruling out running time $\Oh((\mx{X} \sm{X}/(nt))^{1-\eps})$ for any $\eps > 0$ (see Theorem~\ref{thm:seclb_main}).
In particular, for any non-trivial parameter setting with $t \ll \mx{X} \sm{X} / n^2$ this yields a super-linear lower bound. Note that our use of parameter settings ensures that we did not miss any setting in which \SubsetSum admits a near-linear time algorithm.

Our lower bound is conditional on assumptions from fine-grained complexity theory. Specifically, it holds under the Strong Exponential Time Hypothesis~\cite{ip01,cip09}, which is the most standard assumption from fine-grained complexity~\cite{williams2018some} and essentially states that the Satisfiability problem requires time $2^{n-o(n)}$. Alternatively, our lower bound also follows from the Strong k-SUM hypothesis; see Section~\ref{sec:lb:hypo:ksum} for a discussion.
To obtain a uniform lower bound under both of these hypotheses, we introduce the following intermediate hypothesis:

\begin{center}
\emph{For any $\alpha,\eps>0$ there exists $k\ge 3$ such that given a set $Z \subseteq \{1,\ldots,U\}$ of size $|Z| \le U^\alpha$ and given \\ a target $T$, no algorithm decides whether any $k$ numbers in $Z$ sum to $T$ in time $\Oh(U^{1-\eps})$.}
\end{center}

We first show that this intermediate hypothesis is implied both by the Strong Exponential Time Hypothesis (via a reduction from~\cite{AbboudBHS19}) and by the Strong k-SUM hypothesis (which is easy to prove). Then we show that the intermediate hypothesis implies our desired conditional lower bound for every non-trivial parameter setting of \SubsetSum. 
For this step, we design a reduction that starts from a \kSUM instance and constructs an equivalent \SubsetSum instance. This is in principle an easy task. However, here we are in a fine-grained setting, where we cannot afford any polynomial overhead and thus have to be very careful. Specifically, as we want to prove a conditional lower bound for each non-trivial parameter setting, we need to design a family of reductions that is parameterized by $(\tau,\xi,\sigma)$. Ensuring the conditions $t = \Theta(n^\tau)$, $\mx{X} = \Theta(n^\xi)$, and $\sm{X} = \Theta(n^\sigma)$ in the constructed \SubsetSum instance $(X,t)$ requires several ideas on how to ``pack'' numbers (in fact, this task is so complicated that for the case of multi-sets we were not able to prove a tight conditional lower bound).

\section{The Algorithm} \label{sec:algo}

\subsection{Precise Theorem Statements and Combination}

\newcommand{\ThmStructDensity}{\ensuremath{C_\delta}}
\newcommand{\ThmStructAlmost}{\ensuremath{C_\alpha}}
\newcommand{\ThmStructLambda}{\ensuremath{C_\lambda}}

Recall that our algorithm has two main components, the algorithmic part and the structural part. We now present formal statements of these parts.

\begin{restatable}[Algorithmic Part, Formal Version of Theorem~\ref{thm:main_algo}]{theorem}{thmmainalgoformal} \label{thm:main_algo_formal}
  Let $\delta, \alpha$ be functions of $n$ with $\delta \ge 1$ and $16 \alpha \le \delta$.
  Given an $\delta$-dense multiset $X$ of size $n$, in time $\tOh(n)$ we can compute an integer $d \ge 1$ such that $X' := X(d)/d$ is $\delta$-dense and has no $\alpha$-almost divisor. 
  Moreover, we have the following additional properties: 
  \begin{enumerate}
    \item $d \le 4 \mul{X} \sm{X} / \size{X}^2$,
    \item $d = \Oh(n)$,
    \item $|X'| \ge 0.75\, \size{X}$, 
    \item $\sm{X'} \ge 0.75\, \sm{X}/d$.\lipicsEnd
  \end{enumerate}
\end{restatable}

\begin{restatable}[Structural Part, Formal Version of Theorem~\ref{thm:main_struct}]{theorem}{thmmainstructformal} \label{thm:main_struct_formal}
  Let $X$ be a multi-set and set
    \begin{align*}
    C_\delta &:= 1699200 \cdot \log (2n) \log^2(2\mul{X}), \\
    C_\alpha &:= 42480 \cdot \log(2\mul{X}), \\
    C_\lambda &:= 169920\cdot \log(2\mul{X}).
  \end{align*}
  If $X$ is $C_\delta$-dense and has no $C_\alpha$-almost divisor,
  then for $\la{X} := C_\lambda \cdot \mul{X} \mx{X} \sm{X} / |X|^2$ we have 
  \[ \fragment{\la{X}}{\sm{X} - \la{X}} \subseteq
  \ssm{X}.\lipicsEnd \]
\end{restatable}

\medskip
We next show how to combine these theorems to solve \SubsetSum. We use notation as in Theorem~\ref{thm:main_struct_formal}.

\begin{theorem}[Combination I, Formal Version of Theorem~\ref{thm:combine_struct}] \label{thm:combine_struct_formal}
  Given a $\ThmStructDensity$-dense multi-set $X$, in time $\tOh(n)$ we can compute an integer $d \ge 1$ such that for any $t \le \sm{X}/2$ with $t \ge (4+2\ThmStructLambda) \mul{X} \mx{X} \sm{X} / |X|^2$:
  \[ t \in \ssm{X} \quad\text{if and only if}\quad t \bmod d \in \ssm{X} \bmod d. \]
\end{theorem}
\begin{proof}
The easy direction does not depend on the choice of $d$: If $t \bmod d \not\in \ssm{X} \bmod d$, then clearly $t$ is not a subset sum of $X$. 

For the more difficult direction, first apply Theorem~\ref{thm:main_algo_formal} with $\delta := \ThmStructDensity$ and $\alpha := \ThmStructAlmost$ to compute an integer $d \ge 1$ such that $X' = X(d)/d$ has no $C_\alpha$-almost divisor and is $C_\delta$-dense (note that Theorem~\ref{thm:main_algo_formal} is applicable since $16 C_\alpha \le C_\delta$). 
Then
Theorem~\ref{thm:main_struct_formal} is applicable to $X'$ and shows that $\fragment{\lambda_{X'}}{\sm{X'} - \lambda_{X'}} \subseteq \ssm{X'}$. 

Note that if $t \bmod d \in \ssm{X} \bmod d$, then there is a subset $Y \subseteq X$ summing to $t$ modulo $d$.
We choose a minimal such set $Y$, that is, we pick any $Y \subseteq X$ such that $\sm{Y} \bmod d = t \bmod d$, but for any proper subset $Y' \subsetneq Y$ we have $\sm{Y'} \bmod d \ne t \bmod d$. 
We claim that (i) $Y \subseteq \overline{X(d)}$ and (ii) $t' := (t - \sm{Y})/d$ lies in the feasible interval $\fragment{\lambda_{X'}}{\sm{X'} - \lambda_{X'}}$. 
Assuming these two claims, by Theorem~\ref{thm:main_struct_formal} there exists a set $Z' \subseteq X'$ with $\sm{Z'} = t'$. This corresponds to a set $Z \subseteq X(d)$ with $\sm{Z} = d t' = t - \sm{Y}$. Since $Y$ and $Z$ are subsets of disjoint parts of $X$, their union $Y \cup Z$ is a subset of $X$ summing to $t$. This proves the desired statement: if $t \bmod d \in \ssm{X} \bmod d$ then $t \in \ssm{X}$. In the following we prove the two remaining claims.

\begin{claim} \label{cla:wlogY}
  We have (1) $Y \subseteq \overline{X(d)}$, (2) $|Y| \le d$, and (3)
  $ \sm{Y} \,\le\, 4 \mul{X} \mx{X} \sm{X} / \size{X}^2$. 
\end{claim}
\begin{claimproof}
  (1) If $Y \not\subseteq \overline{X(d)}$, then we can replace $Y$ by $Y \cap \overline{X(d)}$ without changing the value of $\sm{Y} \bmod d$, as this change only removes numbers divisible by $d$. Hence, by minimality of $Y$ we have $Y \subseteq \overline{X(d)}$.
  
  To see (2), write $Y = \{y_1,\ldots,y_\ell\}$ and consider the prefix sums $(y_1+\ldots+y_i) \bmod d$. If $\ell > d$, then by the pigeonhole principle there exist $i<j$ with the same remainder 
\[y_1+\ldots+y_i \equiv y_1+\ldots+y_j \pmod d. \] 
It follows that $y_{i+1}+\ldots+y_j \equiv 0 \pmod d$, so we can remove $\{y_{i+1},\ldots,y_j\}$ from $Y$ without changing the value of $\sm{Y} \bmod d$. As this violates the minimality of $Y$, we obtain $\ell \le d$.

  For (3), using Theorem~\ref{thm:main_algo_formal}.1, the inequality $|Y| \le d$ implies $\sm{Y} \,\le\, d \cdot \mx{X} \,\le\, 4 \mul{X} \mx{X} \sm{X} / \size{X}^2$.
\end{claimproof}

The remaining target $t - \sm{Y}$ is divisible by $d$. 
It remains to prove that $t' := (t - \sm{Y})/d$ lies in the feasible interval of Theorem~\ref{thm:main_struct_formal} applied to $X' = X(d)/d$:
\begin{claim}
  We have $t' \in \fragment{\lambda_{X'}}{\sm{X'} - \lambda_{X'}}$.
\end{claim}
\begin{claimproof}
Using the inequality $|X'| \ge 0.75\, |X|$ from Theorem~\ref{thm:main_algo_formal}.3, the bound $(1/0.75)^2 < 2$, and the easy facts $\mul{X'} \le \mul{X}$, $\mx{X'} \le \mx{X}/d$, and $\sm{X'} \le \sm{X}/d$, we bound
\[ \lambda_{X'} \;=\; \frac{\ThmStructLambda \mul{X'} \mx{X'} \sm{X'} }{ |X'|^2 } \;\le\; \frac{ 2 \ThmStructLambda \mul{X} \mx{X} \sm{X} }{ d |X|^2 }.
\]
By the assumption $t \ge (4+2\ThmStructLambda) \mul{X} \mx{X} \sm{X} / |X|^2$ and Claim~\ref{cla:wlogY}.(3), it follows that $t' = (t - \sm{Y})/d \ge \lambda_{X'}$. 

\smallskip
For the other direction, we use that $X$ is $C_\delta$-dense and thus also $8 \ThmStructLambda$-dense, which gives $\mul{X} \mx{X} / |X|^2 \le 1/(8 \ThmStructLambda)$. 
Therefore, we can further bound
\[ \lambda_{X'} \;\le\; \frac{ 2 \ThmStructLambda \mul{X} \mx{X} \sm{X} }{ d |X|^2 } \le \frac{\sm{X}}{4d}.
\]
From Theorem~\ref{thm:main_algo_formal}.4 we have $\sm{X'} \ge 0.75\, \sm{X}/d$, and thus $\sm{X'} - \lambda_{X'} \ge 0.5\, \sm{X}/d$. 
Finally, we use the assumption $t \le \sm{X}/2$ to obtain
\[ t'= \frac{t - \sm{Y}}{d} \le \frac td \le \frac {\sm{X}}{2d} \le \sm{X'} - \lambda_{X'}. \]
This finishes the proof of $t' \in \fragment{\lambda_{X'}}{\sm{X'} - \lambda_{X'}}$.
\end{claimproof}
We thus proved the two remaining claims, finishing the proof.
\end{proof}

This leads to the following formal version of our \emph{final algorithm}. We use notation as in Theorem~\ref{thm:main_struct_formal}.

\begin{theorem}[Combination II, Formal Version of Theorem~\ref{thm:combine_algo}] \label{thm:combine_algo_formal}
  We can preprocess a given $\ThmStructDensity$-dense multi-set~$X$ of size $n$ in time~$\tOh(n)$. Given a query $t \le \sm{X}/2$ with $t \ge (4+2\ThmStructLambda) \mul{X} \mx{X} \sm{X} / n^2$ we can then decide whether $t \in \ssm{X}$ in time $\Oh(1)$.
  
  In particular, given a multi-set $X$ of size $n$ and a target $t \le \sm{X}/2$ with\footnote{Note that by definition of $C_\delta$, the requirement on $t$ is $t \ge 849600 \cdot \log (2n) \log^2(2\mul{X}) \cdot \mul{X} \mx{X} \sm{X} / n^2$. We did not optimize constant factors.} $t \ge 0.5 \ThmStructDensity \mul{X} \mx{X} \sm{X} / n^2$, we can decide whether $t \in \ssm{X}$ in time $\tOh(n)$.
\end{theorem}

\begin{proof}
  In the preprocessing, we first run the algorithm from Theorem~\ref{thm:combine_struct_formal} to compute the number $d$. Then we solve \SubsetSum modulo $d$, that is, we compute the set $\ssm{X} \bmod d$. To this end, we use a recent algorithm by Axiotis et al.~\cite{AxiotisBJTW19,AxiotisBBJNTW20} that runs in time $\tOh(n+d)$. By Theorem~\ref{thm:main_algo_formal}.2 we have $d = \Oh(n)$, so the running time can be bounded by $\tOh(n)$.
  
  On query $t$, we check whether $t \bmod d$ lies in the precomputed set $\ssm{X} \bmod d$. If so, we return ``$t \in \ssm{X}$'', if not, we return ``$t \not\in \ssm{X}$''. This runs in time $\Oh(1)$.
  
  \smallskip
  For the second formulation, note that $\sm{X}/2 \ge t \ge 0.5 \ThmStructDensity \mul{X} \mx{X} \sm{X} / n^2$ implies $n^2 \ge \ThmStructDensity \mul{X} \mx{X}$.
  Hence, the multi-set $X$ is $\ThmStructDensity$-dense. Since also $0.5 \ThmStructDensity \ge 4+2\ThmStructLambda$, the first formulation applies and we can decide whether $t \in \ssm{X}$ in $\tOh(n)$ preprocessing time plus $\Oh(1)$ query time. 
\end{proof}

It remains to prove the two main components: the structural part and the algorithmic part. After some preparations in Section~\ref{sec:algo:preparations}, we will prove the algorithmic part in Section~\ref{sec:algo:algo} and the structural part in Section~\ref{sec:algo:struct}.

\subsection{Preparations}
\label{sec:algo:preparations}

We start with some observations about our notion of density. Recall the following definition.

\defdensity*

\begin{lemma}\label{obs:1}
    For any $\delta$-dense multi-set $X$ of size $n$, we have 
    $\mul{X} \cdot \sm{X} / n^2 \le n/\delta$.
\end{lemma}
\begin{proof}
    By definition of $\delta$-density, we have $n^2 \ge \delta \cdot \mul{X} \cdot \mx{X}$. Combining this with the trivial inequality $\sm{X} \le \mx{X}\cdot n$ and rearranging yields the claim.
\end{proof}
Next we show that large subsets of dense multi-sets are dense as well.
\begin{lemma}\label{obs:2}
    For any $\kappa \ge 1$ and any $\delta$-dense multi-set $X$, any subset $Y \subseteq X$ of size
    $|Y| \ge |X|/\kappa$ is $\delta/\kappa^2$-dense.
\end{lemma}
\begin{proof}
    Using the size assumption $|Y| \ge |X|/\kappa$, the definition of $\delta$-density, and the trivial facts $\mul{Y} \le \mul{X}$ and $\mx{Y} \le \mx{X}$, we obtain \[
    \kappa^2 \cdot |Y|^2 \ge |X|^2
    \ge \delta \cdot \muli{X} \cdot \mxi{X}
    \ge \delta \cdot \muli{Y} \cdot \mxi{Y}.
    \] This yields the claimed inequality after rearranging.
\end{proof}
Lastly, we show that dividing all numbers in a set increases the density of a set.
\begin{lemma}\label{obs:3}
    Let $d \ge 1$ be an integer, and let $X$ be a $\delta$-dense multi-set of positive integers divisible by~$d$. Then the multi-set $X/d$ is $d\delta$-dense.
\end{lemma}
\begin{proof}
    By definition of $\delta$-density, we have\begin{align*}
        |X/d|^2 = |X|^2 &\ge \delta\cdot\muli{X}\cdot \mxi{X}
        = \delta\cdot\muli{X/d}\cdot d \mxi{X/d},
    \end{align*} where we used the facts that the multi-sets $X$ and $X/d$ have the same number of elements and the same multiplicity, while $\mxi{X} = d\cdot \mxi{X/d}$. In total, this proves
    that the multi-set $X/d$ is $d\delta$-dense.
\end{proof}

\subsection{Algorithmic Part} \label{sec:algo:algo}

In this section, we first design an algorithm for prime factorization (proving Theorem~\ref{thm:PF} in Section~\ref{sec:algo:algo:PF}), then use this to find almost divisors (Section~\ref{sec:algo:algo:findad}), and finally present a proof of the algorithmic part (proving Theorem~\ref{thm:main_algo_formal} in Section~\ref{sec:algo:algo:algo}).

\subsubsection{Prime Factorization} \label{sec:algo:algo:PF}

In this section, we show that $n$ given numbers in $\position{s}$ can be factorized in total time $\tOh(n+\sqrt{s})$, proving Theorem~\ref{thm:PF}.
We start by describing the following subroutine. 

\begin{lemma}[Decision Subroutine] \label{lem:subdec}
  Given a set of integers $M \subseteq \position{s}$ and a set of prime numbers $P \subseteq \position{s}$, in time $\tOh((|M|+|P|) \log s)$ we can compute all $m \in M$ that are divisible by some $p \in P$, that is, we can compute the set $M' := \{m \in M \mid \exists p \in P\colon p \text{ divides } m\}$.
\end{lemma}
\begin{proof}
  Observe that an integer $m$ is divisible by a prime $p$ if and only if their greatest common divisor satisfies $\gcd(m,p) > 1$. More generally, $m$ is divisible by some $p \in P$ if and only if we have $\gcd\big( m, \prod_{p \in P} p \big) > 1$. 
  This is the check that we will use in our algorithm. 
  However, note that $\prod_{p \in P} p$ is an $\Omega(|P|)$-bit number, and thus a direct computation of $\gcd\big( m, \prod_{p \in P} p \big)$ requires time $\widetilde{\Omega}(|P|)$. 
  Repeating this operation for all $m \in M$ would require time $\widetilde{\Omega}(|M| \cdot |P|)$, which we want to avoid.
  
  In the following we make use of efficient algorithms for multiplication and division with remainder, that is, we use that the usual arithmetic operations on $b$-bit numbers take time $\tOh(b)$.
  
  \newcommand{\leftchild}{\textup{leftchild}}
  \newcommand{\rightchild}{\textup{rightchild}}
  \newcommand{\parent}{\textup{parent}}
  For a set $S = \{s_1,\ldots,s_n\}$, we denote by $T_S$ a balanced binary tree with $n$ leaves corresponding to the elements $s_1,\ldots,s_n$. 
  Each node of~$T_S$ corresponds to a subset $I = \{s_i,\ldots,s_j\} \subseteq S$. 
  The root of $T_S$ corresponds to the set $S$. 
  A node corresponding to a set $I \subseteq S$ of size $|I|=1$ is a \emph{leaf} and has no children. A node corresponding to a set $I \subseteq S$ of size $|I|>1$ is an \emph{internal} node and has two children, corresponding to the two parts of a balanced partitioning $I = I_1 \cup I_2$. We denote these children by $\leftchild(I) = I_1$ and $\rightchild(I) = I_2$. Moreover, we denote the parent relation by $\parent(I_1) = \parent(I_2) = I$. We write $V(T_S)$ for the family of all subsets $I$ forming the nodes of~$T_S$.
  
  We construct the trees $T_M$ and $T_P$ in time $\tOh(|M| + |P|)$. 
  
  For each $I \in V(T_M)$ we define $\Pi(I) := \prod_{m \in I} m$. In particular, at any leaf $I = \{m\}$ we have $\Pi(I) = m$, and at the root we have $\Pi(M) = \prod_{m \in M} m$. 
  We compute the numbers $\Pi(I)$ by traversing the tree $T_M$ \emph{bottom-up}, that is, for any internal node $I$ we compute $\Pi(I) := \Pi(\leftchild(I)) \cdot \Pi(\rightchild(I))$. 
  
  To analyze the running time to compute all numbers $\Pi(I)$, note that the total bit length of the numbers $\Pi(I)$ on any fixed level of $T_M$ is $\Oh(|M| \log s)$. Using an efficient multiplication algorithm, we can thus perform all operations on a fixed level in total time $\tOh(|M| \log s)$. Over all $\Oh(\log |M|)$ levels of $T_M$, the running time is still bounded by $\tOh(|M| \log s)$.
  
  In the same way, we compute the numbers $\Pi(J) = \prod_{p \in J} p$ for all nodes $J \in V(T_P)$. In particular, at the root of $T_P$ we compute $\Pi(P) = \prod_{p \in P} p$. This takes time $\tOh(|P| \log s)$.
  
  Now we compute the number $\Pi(P) \bmod \Pi(M)$. Since the combined bit length of $\Pi(P)$ and $\Pi(M)$ is $\Oh((|M|+|P|) \log s)$, this takes time $\tOh((|M|+|P|) \log s)$.
  
  Next we compute the numbers $R(I) := \Pi(P) \bmod \Pi(I)$ for all $I \in V(T_M)$, by traversing the tree $T_M$ \emph{top-down}. At the root we use the already computed number $\Pi(P) \bmod \Pi(M)$. At an internal node $I$ we compute 
  \[ R(I) := R(\parent(I)) \bmod \Pi(I). \]
  Note that both $R(\parent(I))$ and $\Pi(I)$ are already computed when we evaluate $R(I)$.
  
  To analyze the running time to compute all numbers $R(I)$, we note that $R(I) \le \Pi(I)$, so the total bit length of the numbers $R(I)$ is bounded by the total bit length of the numbers $\Pi(I)$. Thus, the same analysis as before shows that computing the numbers $R(I)$ takes total time $\tOh(|M| \log s)$.
  
  Note that for the leaves of $T_M$ we have now computed the numbers $R(\{m\}) = \Pi(P) \bmod m$ for all $m \in M$.
  Since $m$ and $R(\{m\})$ have bit length $\Oh(\log s)$, we can compute their greatest common divisor $\gcd(m, R(\{m\}))$ in time $\tOh(\log s)$. In total, this takes time $\tOh(|M| \log s)$.
  
  Finally, we use the identity $\gcd(a,b) = \gcd(a, b \bmod a)$ to observe that 
  \[ \gcd\Big(m, \prod_{p \in P} p\Big) = \gcd\big(m, \Pi(P)\big) = \gcd\big(m, R(\{m\})\big). \]
  By our initial discussion, a number $m \in M$ is divisible by some $p \in P$ if and only if $\gcd(m, \prod_{p \in P} p) > 1$. Hence, we can determine all $m \in M$ that are divisible by some $p \in P$ in total time $\tOh((|M|+|P|) \log s)$.
\end{proof}

Next we adapt the above decision subroutine to obtain a search subroutine.

\begin{lemma}[Search Subroutine] \label{lem:subsearch}
  Given a set of integers $M \subseteq \position{s}$ and a set of prime numbers $P \subseteq \position{s}$, in time $\tOh((|M|+|P|) \log^2 s)$ we can compute for all numbers $m \in M$ all prime factors among the primes $P$, that is, we can compute the set $F := \{(m,p) \mid m \in M,\, p \in P, \, p \text{ divides } m\}$.
\end{lemma}
\begin{proof}
  On instance $(M,P)$, we first partition in a balanced way $P = P_1 \cup P_2$. 
  Then we call the decision subroutine from Lemma~\ref{lem:subdec} twice, to compute the sets $M_1$ and $M_2$ with
  \begin{align*}
    M_i &\,:=\, \{m \in M \mid \exists p \in P_i\colon p \text{ divides } m\} \qquad \text{for }\, i \in \{1,2\}.
  \end{align*}
  Finally, we recursively solve the instances $(M_1,P_1)$ and $(M_2,P_2)$. (We ignore recursive calls with $M = \emptyset$.)
  
  In the base case, we have $P = \{p\}$. In this case, all $m \in M$ are divisible by $p$, so we print the pairs $\{(m,p) \mid m \in M\}$.
  
  \smallskip
  Correctness of this algorithm is immediate. 
  To analyze its running time, note that the recursion depth is $\Oh(\log |P|)$, since $P$ is split in a balanced way. Further, the total size of $P'$ over all recursive calls $(M',P')$ on a fixed level of recursion is $\Oh(|P|)$. 
  Moreover, note that a number $m \in \position{s}$ has $\Oh(\log s)$ prime factors. Since for each recursive call $(M',P')$ each number $m \in M'$ has a prime factor in $P'$, it follows that each number $m \in M$ appears in $\Oh(\log s)$ recursive calls on a fixed level of recursion. Thus, the total size of $M'$ over all recursive calls $(M',P')$ on a fixed level of recursion is $\Oh(|M| \log s)$. 
  Plugging these bounds into the running time $\tOh((|M|+|P|) \log s)$ of Lemma~\ref{lem:subdec} yields a total time of $\tOh((|M| \log s + |P|) \log s) = \tOh((|M|+|P|) \log^2 s)$ per level. The same time bound also holds in total over all $\Oh(\log |P|)$ levels.
\end{proof}

Our main prime factorization algorithm now follows easily.

\thmPF*

\begin{proof}
  Let $M \subseteq \position{s}$ be the set of $n$ given numbers.
  Denote by $P$ the set of all prime numbers less than or equal to $\sqrt{s}$, and note that $P$ can be computed in time $\tOh(\sqrt{s})$, e.g., by using the Sieve of Eratosthenes. 
  We run Lemma~\ref{lem:subsearch} on $(M,P)$ to obtain the set $F := \{(m,p) \mid m \in M,\, p \in P, \, p \text{ divides } m\}$.
  From this set, we can compute the prime factorization of each $m \in M$ efficiently as follows.
  Fix $m \in M$. For any $(m,p) \in F$, determine the largest exponent $e = e(m,p)$ such that $p^e$ divides $m$. 
  This determines all prime factors of $m$ that are less than or equal to $\sqrt{s}$. Since $m \in \position{s}$, the number $m$ has at most one prime factor greater than $\sqrt{s}$. We determine this potentially missing prime factor as $q := m / \prod_{p \in P} p^{e(m,p)}$. If $q > 1$, then $q$ is the single large prime factor of $m$, and if $q=1$, then all prime factors of $m$ are less than or equal to $\sqrt{s}$.
  The running time is dominated by the call to Lemma~\ref{lem:subsearch}, which takes time 
  \[ \tOh((|M|+|P|)\log^2 s) = \tOh((n+\sqrt{s})\log^2 s) = \tOh(n+\sqrt{s}). \qedhere \]
\end{proof}

\subsubsection{Finding an Almost Divisor} \label{sec:algo:algo:findad}

In this section, we use the prime factorization algorithm from the last section to find almost divisors.

\begin{theorem}[Finding Almost Divisors] \label{thm:findad}
  Given $\alpha > 0$ and a multi-set $X$ of size $n$, we can decide whether $X$ has an $\alpha$-almost divisor, and compute an $\alpha$-almost divisor if it exists, in time $\tOh(n+\sqrt{\mx{X}})$.
\end{theorem}

Recall the definition of almost divisors:

\defalmostdivisor*

We first observe that any proper divisor $d'$ of an almost divisor $d$ is also an almost divisor.

\begin{lemma}\label{ft:pad}
    If $d$ is an $\alpha$-almost divisor of a multi-set $X$, then any divisor $d' > 1$ of $d$ is also an $\alpha$-almost divisor of $X$.
\end{lemma}
\begin{proof}
    Since any number divisible by $d$ is also divisible by $d'$, we have $|X(d')| \ge |X(d)|$, or, equivalently, $|\overline{X(d')}| \le |\overline{X(d)}|$. By definition of $\alpha$-almost divisor, we obtain 
    \[ |\overline{X(d')}| \le |\overline{X(d)}| \le \alpha \mul{X} \sm{X} / |X|^2, \]
    so also $d'$ is an $\alpha$-almost divisor. 
\end{proof}

The above lemma shows that if $X$ has an $\alpha$-almost divisor, then it also has a \emph{prime} $\alpha$-almost divisor, that is, it has an $\alpha$-almost divisor that is a prime number.
This suggests the following approach.

\begin{proof}[Proof of Theorem~\ref{thm:findad}]
 We use Theorem~\ref{thm:PF} to compute the prime factorization of all numbers in $X$ in total time $\tOh(n + \sqrt{\mx{X}})$. 
From the prime factorizations we can infer for each prime $p$ (that divides some $x \in X$) the number of $x \in X$ that are divisible by $p$. In particular, we can determine whether some prime~$p$ divides at least $n - \alpha \mul{X} \sm{X} / n^2$ numbers in $X$. If such a prime $p$ exists, then $p$ is an $\alpha$-almost divisor of $X$. If no such prime $p$ exists, then the set $X$ has no prime $\alpha$-almost divisor, so by \cref{ft:pad} the set $X$ has no $\alpha$-almost divisor. 
This proves Theorem~\ref{thm:findad}.
\end{proof}

\subsubsection{Proof of the Algorithmic Part} \label{sec:algo:algo:algo}

We are now ready to prove our algorithmic part.

\thmmainalgoformal*

\begin{algorithm}[t]
    \SetKwBlock{Begin}{}{end}
    \SetAlgoNoLine
    \SetKwFunction{aldiv}{AlmostDivisorFreeSubset}

    \aldiv{$\alpha$, $X$}\Begin{
        $X_0 \gets X$; $i \gets 1$\;
        \While{$X_{i-1}$ has an $\alpha$-almost divisor $d_i$}{
            $X_i \gets X_{i-1}(d_i)/d_i$\;
            $i\gets i + 1$\;
        }
        $d \gets d_1\cdots d_i$\;
        \Return{$(d, X_i)$};
    }
    \caption{Reduction to the almost-divisor-free setting, see Theorem~\ref{thm:main_algo_formal}.}\label{alg:aldiv}
\end{algorithm}

\medskip
Consider \cref{alg:aldiv}, which iterative removes almost divisors. We start with $X_0 = X$. While $X_{i-1}$ has an $\alpha$-almost divisor $d_i$, we continue with $X_i := X_{i-1}(d_i)/d_i$, that is, we remove all numbers not divisible by $d_i$ from $X_{i-1}$ and divide the remaining numbers by $d_i$.
The final multi-set $X_i$ has no $\alpha$-almost divisor. We return $d = d_1 \cdots d_i$.

\medskip
We first analyze the running time of this algorithm. 
By Theorem~\ref{thm:findad}, we can find almost divisors in time $\tOh(n + \sqrt{\mx{X}})$. This dominates the running time of one iteration of \cref{alg:aldiv}.
Note that the number of iterations is bounded by $\Oh(\log \mx{X})$, since  $\mx{X_i} = \mx{X} / (d_1 \cdots d_i)$ and $d_1 \cdots d_i \ge 2^i$. Therefore, the total running time of \cref{alg:aldiv} is $\tOh(n + \sqrt{\mx{X}})$. Rearranging the definition of $\delta$-density yields $\mx{X} \le n^2 / (\delta \mul{X})$. Since $\mul{X} \ge 1$ and $X$ is $\delta$-dense for $\delta \ge 1$, we obtain $\mx{X} = \Oh(n^2)$. Hence, the running time is $\tOh(n + \sqrt{\mx{X}}) = \tOh(n)$, as claimed in Theorem~\ref{thm:main_algo_formal}.

\medskip
In the following we analyze correctness of \cref{alg:aldiv}, that is, we show that it ensures the properties claimed in Theorem~\ref{thm:main_algo_formal}. 
We will denote by $X_i$ any intermediate multi-set of \cref{alg:aldiv}, for $i \ge 0$.

Observe that $X_i$ contains all numbers in $X$ that are divisible by $d_1 \cdots d_i$, divided by $d_1 \cdots d_i$. That is, 
\[ X_i = X(d_1 \cdots d_i) / (d_1 \cdots d_i). \]
In particular, \cref{alg:aldiv} returns the multi-set $X(d)/d$.
Since $X(d_1 \cdots d_i) \subseteq X$, we obtain the easy facts
\begin{align}
  \mx{X_i} &\le \mx{X} / (d_1 \cdots d_i), \notag \\
  \sm{X_i} &\le \sm{X} / (d_1 \cdots d_i), \label{eq:easyfacts} \\
  \mul{X_i} &\le \mul{X}. \notag 
\end{align}
The key property in our analysis is the size $|X_i|$, and how it compares to $n = |X|$.
\begin{claim} \label{cla:sizeXi}
  For any $i \ge 0$, we have
  \[
    |X_i| \;\ge\; n - \frac{ 4\alpha \cdot \mul{X}\sm{X} }{ n^2 } \;\ge\; \Big(1 - \frac{4\alpha}\delta\Big) n \;\ge\; \frac 34 n.
  \]
\end{claim}
\begin{proof}
  Since $d_{i+1}$ is an $\alpha$-almost divisor of $X_i$, at most $\alpha \cdot \mul{X_i} \sm{X_i} / |X_i|^2$ numbers in $X_i$ are not divisible by $d_i$. For $X_{i+1} = X_i(d_{i+1})/d_{i+1}$ we can thus bound
  \[ |X_{i+1}| = |X_i(d_{i+1})| \ge |X_i| - \frac{\alpha \cdot \mul{X_i} \sm{X_i} }{ |X_i|^2 }. \]
  Using the easy facts (\ref{eq:easyfacts}), we obtain
  \begin{align}
    |X_{i+1}| \ge |X_i| - \frac{ \alpha \cdot \mul{X} \sm{X} }{ d_1 \ldots d_i |X_i|^2 } \ge |X_i| - \frac{ \alpha \cdot \mul{X} \sm{X} }{ 2^i |X_i|^2 }. \label{eq:recurrence}
  \end{align}
  We use this inequality to inductively prove that
  \begin{align}
    |X_i| \ge n - \Big(1 - \frac 1{2^i}\Big) \frac{4\alpha \cdot \mul{X} \sm{X}}{n^2}. \label{eq:inductivehypo}
  \end{align}
  Let us first argue that this inequality implies the main claim.
  Using $1 - 1/2^i \le 1$, we obtain the first claimed inequality
  \[ |X_i| \ge n - \frac{4\alpha \cdot \mul{X} \sm{X}}{n^2}. \]
  Since $X$ is $\delta$-dense, Observation~\ref{obs:1} yields $\mul{X} \sm{X} \le n^3 / \delta$. Plugging this in, we obtain the second inequality
  \[ |X_i| \ge n \cdot \Big(1 - \frac{4\alpha}{\delta} \Big). \]
  The last inequality $|X_i| \ge \frac 34 n$ now follows from the assumption $16 \alpha \le \delta$ of Theorem~\ref{thm:main_algo_formal}.
  
  It remains to prove inequality (\ref{eq:inductivehypo}) by induction.
  The inductive base is $i = 0$ with $X_0 = X$ and thus $|X_0| = n$. 
  For the inductive step, assume that the induction hypothesis (\ref{eq:inductivehypo}) holds for $X_i$. 
  As shown above, the inductive hypothesis for $X_i$ implies $|X_i| \ge \frac 34 n$. 
  Plugging this bound into the recurrence (\ref{eq:recurrence}) yields
  \[ |X_{i+1}| \ge |X_i| - \frac{ (4/3)^2 \alpha \cdot \mul{X} \sm{X} }{ 2^i n^2 } \ge |X_i| - \frac{ 4 \alpha \cdot \mul{X} \sm{X} }{ 2^{i+1} n^2 }. \]
  Using the induction hypothesis (\ref{eq:inductivehypo}) again, we obtain
  \[ |X_{i+1}| \ge n - \Big(1 - \frac 1{2^i}\Big) \frac{4\alpha \cdot \mul{X} \sm{X}}{n^2} - \frac{ 4 \alpha \cdot \mul{X} \sm{X} }{ 2^{i+1} n^2 } = n - \Big(1 - \frac 1{2^{i+1}}\Big) \frac{4\alpha \cdot \mul{X} \sm{X}}{n^2}. \]
  This finishes the inductive step, and thus the proof of the claim.
\end{proof}

The claimed properties of the multi-set $X' := X(d)/d$ computed by \cref{alg:aldiv} now easily follow from Claim~\ref{cla:sizeXi}, as we show in the following.

\begin{claim}
  $X'$ is $\delta$-dense.
\end{claim}
\begin{claimproof}
  By Claim~\ref{cla:sizeXi} we have $|X(d)| = |X'| \ge \frac 34 |X|$. Since $X$ is $\delta$-dense, Observation~\ref{obs:2} implies that $X(d) \subseteq X$ is $(\frac 34)^2 \delta$-dense; in particular it is $\delta/2$-dense. Now Observation~\ref{obs:3} implies that $X' = X(d)/d$ is $d \delta/2$-dense. If $d > 1$ then $d \delta /2 \ge \delta$, so $X'$ is $\delta$-dense. If $d=1$, then $X'=X$, which is $\delta$-dense.
\end{claimproof}

This finishes the proof of the main statement of Theorem~\ref{thm:main_algo_formal}. It remains to verify the four additional properties.

\begin{claim}
  We have $|X'| \ge 0.75\, n$.
\end{claim}
\begin{claimproof}
  Follows directly from Claim~\ref{cla:sizeXi}.
\end{claimproof}

\begin{claim}
  We have $\sm{X'} \ge 0.75\, \sm{X}/d$.
\end{claim}
\begin{claimproof}
  We can bound the sum of the removed elements by 
  \[ \sm{\overline{X(d)}} \le |\overline{X(d)}| \cdot \mx{X} \le \frac{ 4\alpha \cdot \mul{X}\sm{X} }{ |X|^2 } \cdot \mx{X}, \]
  where we used the first inequality of Claim~\ref{cla:sizeXi}. Using that $X$ is $\delta$-dense, we obtain
  \[ \sm{\overline{X(d)}} \le \frac{ 4\alpha  }{ \delta } \sm{X}. \]
  By the assumption $16 \alpha \le \delta$ from Theorem~\ref{thm:main_algo_formal}, we obtain $\sm{\overline{X(d)}} \le \sm{X}/4$. 
  Finally, we note that 
  \[ \sm{X'} = \sm{X(d)/d} = \frac 1d \big(\sm{X} - \sm{\overline{X(d)}}\big) \ge \frac{3 \sm{X}}{4d}. \qedhere \]
\end{claimproof}

\begin{claim}[Compare {\cite[Lemma 3.10]{gm91}}]
  We have $d \le 4 \mul{X} \sm{X} / n^2$.
\end{claim}
\begin{claimproof}
    \newcommand*\dif{\mathop{}\!\mathrm{d}}
  Sort the numbers in $X(d) = \{x_1 \le \dots \le x_{|X(d)|}\}$ and define a
    function $f(z) := x_{\lceil z\rceil}$. Note that, as the numbers in $X(d)$ are
    divisible by $d$ and each number appears at most $\mul{X}$ times in $X(d)$, we can lower bound the value of the function $f$ at $z$ by
    $f(z) = x_{\lceil z \rceil} \ge \lceil\lceil z\rceil / \muli{X}\rceil \cdot d \ge zd/\muli{X}$.
    Now, we can write the sum $\sm{X(d)}$ as the integral of the function $f$
    from $0$ to $|X(d)|$:\[
        \sm{X} \ge \sm{X(d)} = \!\int\limits_0^{|X(d)|}\!\!\!f(z) \dif z \,\ge\,
        \frac{d}{\mul{X}}\,\cdot\!\!\int\limits_0^{|X(d)|}\!\!\,z \dif z
        \,=\,\frac{d}{\mul{X}}\cdot \frac{{|X(d)|}^2}{2}.
    \] 
    Using $|X(d)| = |X'| \ge \frac 34 n$ from Claim~\ref{cla:sizeXi}, we obtain
    \[ \sm{X} \,\ge\, \frac d{\mul{X}} \cdot \frac {n^2}{(4/3)^2\cdot 2} \,\ge\, \frac{d\, n^2}{4 \mul{X}}. \]
    Rearranging now yields the claim.
\end{claimproof}

\begin{claim}
  We have $d = \Oh(n)$.
\end{claim}
\begin{claimproof}
  In the preceeding claim we showed that $d \le 4 \mul{X} \sm{X} / n^2$. Using the trivial inequality $\sm{X} \le n \cdot \mx{X}$, we obtain $d \le 4 \mul{X} \mx{X} / n$. Using that $X$ is $\delta$-dense for $\delta \ge 1$, we now obtain $d \le 4 n / \delta = \Oh(n)$.
\end{claimproof}

The above claims verify all claimed properties and thus finish the proof of Theorem~\ref{thm:main_algo_formal}.

\subsection{Structural Part} \label{sec:algo:struct}

In this section, we first show how to find a small subset $R \subseteq X$ that generates all remainders modulo all small numbers $d$ (Section~\ref{sec:genremainder}). Then we construct long arithmetic progressions (Section~\ref{sec:AP}). 
We use these tools to obtain a decomposition of $X$ (Section~\ref{sec:decomposition}), which then yields the structural part (proving Theorem~\ref{thm:main_struct_formal} in Section~\ref{sec:struct_proof}).

Throughout this section, for multi-sets $X,Y$ we write $X + Y$ to denote their \emph{sumset} (the sumset is a set, that is, each distinct sum appears only once in the sumset):
\[X + Y := \{ x + y \mid x\in X, y\in Y\}.\] Further, we write
$X^{h} := X + \dots + X$ for the iterated sumset containing all sums of $h$ (not necessarily distinct) elements of $X$. Similarly, we write $X^{\le h} := \bigcup_{j=1}^{\lfloor h \rfloor} X^j$. Note that the objects $X+Y, X^h$, and~$X^{\le h}$ are sets, not multi-sets.

\subsubsection{Generating All Remainders} \label{sec:genremainder}

\begin{theorem}[{Compare \cite[Theorem 3.4]{gm91}}]\label{lm:t34}
    Let $\delta, \alpha \ge 1$.
    Let $X$ be a $\delta$-dense multi-set of size $n$ that has no $\alpha$-almost divisor. Then there exists a subset $R \subseteq X$ such that 
    \begin{itemize}
      \item $|R| \le |X| \cdot 8\alpha \log (2n) / \delta$,
      \item $\sm{R} \le \sm{X} \cdot 8\alpha \log (2n) / \delta$, and
      \item for any integer $1 \,<\, d \,\le\, \alpha \cdot \mul{X} \sm{X} / n^2$ the multi-set $R$ contains at least $d$ numbers not divisible by~$d$, that is, $|\overline{R(d)}| \ge d$.
    \end{itemize}
\end{theorem}

\begin{proof}
  If $8\alpha \log (2n) \ge \delta$ then we can simply set $R = X$. The first two claims hold trivially, and the third claim holds because $X$ has no $\alpha$-almost divisor, which implies $|\overline{X(d)}| \ge \alpha \cdot \mul{X} \sm{X} / n^2 \ge d$.
  Therefore, from now on we can assume $8\alpha \log (2n) < \delta$. In particular, we have 
  \begin{align}
    \delta > 8 \alpha. \label{eq:sfosrraf}
  \end{align}
  
  Set $\tau := \lceil \alpha \cdot \mul{X} \sm{X} / n^2 \rceil$. By $\alpha \ge 1$ and the easy fact $\sm{X} \ge \sum_{i=1}^n \lceil i / \mul{X} \rceil \ge \frac12 n^2/\mul{X}$ we have $\alpha \cdot \mul{X} \sm{X} / n^2 \ge 1/2$ and thus 
  \[ \alpha \cdot \mul{X} \sm{X} / n^2 \;\le\; \tau \;\le\; 2 \alpha \cdot \mul{X} \sm{X} / n^2. \]
  Since $X$ is $\delta$-dense, Observation~\ref{obs:1} and inequality (\ref{eq:sfosrraf}) now imply that
  \[ \tau \;\le\; 2 \alpha \cdot n / \delta \;<\; n/2. \]
  
  \smallskip
  We start by picking an arbitrary subset $R' \subseteq X$ of size $2\tau$. This is possible because $\tau < n/2$.
  \begin{claim}
    Let $P$ be the set of primes $p$ with $p \le \tau$ and $|\overline{R'(p)}| < \tau$.
    Then we have $|P| \le 2 \log \mx{X}$.
  \end{claim}
  \begin{claimproof}
    Consider the prime factorization of the numbers in $R'$, that is, consider the set 
    \[ \PF{R'} = \{(r,p) \mid r \in R',\, \text{prime $p$ divides $r$}\}. \]
    Since any integer $m \ge 1$ has at most $\log m$ prime factors, we have $|\PF{R'}| \le |R'| \cdot \log \mx{X}$. 
    On the other hand, for each $p \in P$ we have $|R'(p)| \ge \tau = |R'|/2$, so $\PF{R'}$ contains at least $|R'|/2$ pairs of the form~$(r,p)$. Hence, we have
    \[ |P| \cdot \frac{|R'|}2 \le |\PF{R'}| \le |R'| \log \mx{X}, \]
    which yields the claimed bound $|P| \le 2 \log \mx{X}$.
  \end{claimproof}
  For any $p \in P$, we let $R_p \subseteq \overline{X(p)}$ be an arbitrary subset of size $\tau$. This exists by the assumption that $X$ has no $\alpha$-almost divisor.
  
  Finally, we construct the multi-set $R \subseteq X$ as
  \[ R := R' \cup \bigcup_{p \in P} R_p. \]
  (To be precise, we set $\mu(x;R) := \max\{\mu(x;R'), \max\{\mu(x;R_p) \mid p \in P\}\}$, ensuring that $R$ is a subset of~$X$.)
  
  \medskip
  We show that $R$ satisfies the claimed properties. For the third property, consider any integer $1 < d \le \tau$, and let $p$ be any prime factor of $d$. Note that we have $|\overline{R(d)}| \ge |\overline{R(p)}|$, since any number divisible by $d$ is also divisible by $p$. If $p \in P$, then we obtain 
  \[ |\overline{R(p)}| \ge |\overline{R_p(p)}| = |R_p| = \tau \ge d. \]
  If $p \not\in P$, then by construction of $P$ we have
  \[ |\overline{R(p)}| \ge |\overline{R'(p)}| \ge \tau \ge d. \]
  In any case, we have $|\overline{R(d)}| \ge d$, so we proved the third claim.
  
  For the first two claims, note that 
  \[ |R| \;\le\; |R'| + \sum_{p \in P} |R_p| \;=\; 2\tau + |P| \cdot \tau \;\le\; 2(1 + \log \mx{X}) \tau \;\le\; 4 \log (2\mx{X}) \cdot \alpha \cdot \mul{X} \sm{X} / n^2. \]
  Since $X$ is $\delta$-dense for $\delta \ge 1$, we have $\mx{X} \le n^2$, so we can further bound
  \[ |R| \;\le\; 8 \log (2n) \cdot \alpha \cdot \mul{X} \sm{X} / n^2. \]
  Moreover, Observation~\ref{obs:1} now yields
  \[ |R| \;\le\; 8\log (2n) \cdot \alpha \cdot n / \delta, \]
  proving the first claim.
  We similarly bound $\sm{R}$ by
  \[ \sm{R} \;\le\; |R| \cdot \mx{R} \;\le\; 8\log (2n) \cdot \alpha \cdot \mul{X} \mx{X} \sm{X} / n^2. \]
  Using that $X$ is $\delta$-dense, we finally obtain the second claim
  \[ \sm{R} \;\le\; \frac{8 \alpha \log (2n) }{\delta} \sm{X}. \qedhere \]
\end{proof}

Next we show that the set $R$ constructed in the above Lemma~\ref{lm:t34} generates all remainders modulo any small integer $d$. More precisely, with notation as in Lemma~\ref{lm:t34}, the following theorem implies that $\ssm{R}\!\bmod{d} = \mathbb{Z}_d$ holds for any $1 < d \le \alpha \cdot \mul{X} \sm{X} / |X|^2$.  

\begin{theorem}[Compare {\cite[Lemma 3.3]{gm91}}]\label{lm:naco}
    Let $X$ be a multi-set and let $\tau$ be an integer.
    Suppose that for any $1 < d \le \tau$ the multi-set $X$ contains $d$
    numbers not divisible by $d$, that is, $|\overline{X(d)}| \ge d$.
    Then for any $1 \le d \le \tau$
    the set $\ssm{X}$ is \emph{$d$-complete}, that is, \(\ssm{X}\!\bmod{d} = \mathbb{Z}_d\).
\end{theorem}
\begin{proof}
    We perform induction on $d$.
    For $d = 1$ the statement is trivial.

    So consider a number $1 < d \le \tau$. 
    By assumption we have we have $|\overline{X(d)}| \ge d$. We denote the elements of the multi-set $\overline{X(d)}$ by $x_1,\ldots,x_r$, where $r = |\overline{X(d)}| \ge d$.
    
    Let $C_i := \ssm{\{x_1,\dots, x_i\}}\!\bmod{d}$ denote the set of remainders
    that can be obtained from the elements $x_1, \dots, x_i$.
    In other words, we construct the following sequence of sets:\begin{align*}
        C_0 &:= \{ 0 \},\\
        C_i &:= (C_{i-1} + \{0, x_i\}) \bmod d.
    \end{align*} 
    Observe that we have $\ssm{X}\!\bmod{d} = \ssm{\overline{X(d)}} \!\bmod d = C_r$, since numbers divisible by $d$ do not yield new remainders modulo $d$.
    Furthermore, we have 
    \[ 1 = |C_0| \le |C_1| \le \ldots \le |C_r| \le d. \]
    Since $r \ge d$, by the pigeonhole principle we have $|C_{i-1}| = |C_i|$ for some $i$. 
    For this $i$, for any number $c \in C_{i-1}$ also the number $(c+x_i)\bmod d$
    is contained in the set~$C_{i-1}$. 
    More generally, for any positive integer $k$ also the number
    $(c+kx_i)\bmod d$ is contained in the set~$C_{i-1}$. Now we use that the numbers $k x_i \bmod d$ form the subgroup $g \mathbb{Z}_{d/g}$ of $\mathbb{Z}_d$, where $g := \gcd(x_i,d)$. It thus follows that for any $c \in C_{i-1}$ and any integer $k$ also the number $(c+kg) \bmod d$ is in $C_{i-1}$. We call this property \emph{$g$-symmetry}.
    
    Note that we can write
    \[ \ssm{X}\!\bmod d = (C_{i-1} + \ssm{\{x_i,\ldots,x_r\}})\bmod d. \]
    From this, we see that the $g$-symmetry of $C_{i-1}$ extends to $\ssm{X}\!\bmod d$. More precisely, for any $c \in \ssm{X}\!\bmod d$ and any integer $k$, also $(c + kg)\bmod d$ is in $\ssm{X}\!\bmod d$. 
    
    Moreover, since $x_i \in \overline{X(d)}$ is not divisible by $d$, we have $g = \gcd(x_i,d) < d$. Therefore, by induction hypothesis $\ssm{X}$ is $g$-complete. 
    
    Combining $g$-symmetry and $g$-completeness proves that $\ssm{X}$ is $d$-complete. Indeed, for any remainder $z \in \mathbb{Z}_d$, since $\ssm{X}$ is $g$-complete there is a subset sum $y \in \ssm{X}$ with $y \equiv z \pmod g$. Equivalently, we can write $z-y = kg$ for some integer $k$. Since $y \bmod d$ is in $\ssm{X}\!\bmod d$, by the $g$-symmetry property also $(y+kg)\bmod d = z \bmod d$ is in $\ssm{X}\!\bmod d$. Since $z$ was arbitrary, the set $\ssm{X}$ is $d$-complete.
\end{proof}

\subsubsection{Long Arithmetic Progressions} \label{sec:AP}

In this paper, an \emph{arithmetic progression} is a set $\mathcal{P}$ of the form $\{a+s,a+2s,\ldots,a+m\,s\}$. We call $m$ the \emph{length} of $\mathcal{P}$ and $s$ the \emph{step size} of $\mathcal{P}$. 

Proving existence of a long arithmetic progression in a set $\ssm{X}$ has a long tradition, e.g., consider the following result by S{\'a}rk{\"o}zy \cite{s94} (more precisely, we present a variant with improved constants from \cite{l03}).
\begin{theorem}[{{\cite{s94,l03}}}]\label{tm:aps}
    Let $X$ be a set of $n$ positive integers.

    For every integer $4\mx{X} \le m \le n^2 / (12\log(4\mx{X}/n))$
    the set $\ssm{X}$ contains an arithmetic
    progression $\mathcal{P}$ of length $m$ and step size $s \le 4 \mx{X}/n$. Moreover, every element of  $\mathcal{P}$ can be obtained
    as the sum of at most $6 m / n$ distinct elements of $X$.\lipicsEnd
\end{theorem}
Note that this theorem assumes $X$ to be a set.
Unfortunately, such a result is not readily available for multi-sets with prescribed multiplicity $\mul{X}$. 

We remark that one could naively use Theorem~\ref{tm:aps} on multi-sets by ignoring the multiplicities and working on the support $\supp{X}$. However, this loses a factor of $|X| / |\supp{X}| \le \mul{X}$ in the size, and thus changes the density. In particular, this approach would require us to start with an $\Omega(\mul{X})$-dense multi-set~$X$.
We will avoid this additional factor $\mul{X}$, and only pay factors of the form $\textup{polylog}(\mul{X})$.

\medskip
The main result of this section is a theorem similar to Theorem~\ref{tm:aps} that works for $\Omega(\log(n) \log^2(\muli{X}))$-dense multi-sets. 
We prove this result by following and suitably adapting the proof by S{\'a}rk{\"o}zy~\cite{s94}.

\begin{restatable}{theorem}{tmap} \label{tm:ap}
    Let $X$ be a multi-set of size $n$.
    For every integer $m$ with
    \[ 2\mx{X} \le m \le \frac{n^2}{33984 \mul{X} \log (2n) \log^2(2\mul{X})}, \]
    the set $\ssm{X}$
    contains an arithmetic progression~$\mathcal{P}$ of length $m$ and step size 
    $s \le 4248 \mul{X} \mx{X} \log(2\mul{X})/n$.
    Moreover, every element of $\mathcal{P}$
    can be obtained as the sum of at most $4248 \,m\, \mul{X} \log(2\mul{X}) / n$ 
    distinct\footnote{Here, \emph{distinct} means that any integer $x \in X$ may
        be chosen up to its multiplicity $\mu(x;X)$ times. Thus, the elements chosen from $X$ are distinct,
    but the corresponding integers might not.}
    elements of~$X$, and we have $\mx{\mathcal{P}} \le 4248 m\, \mul{X} \mx{X} \log(2\mul{X})/n$.\lipicsEnd
\end{restatable}

The proof of Theorem~\ref{tm:ap} proceeds similar as in \cite{s94}; we present it here for completeness.
Similar to~\cite{s94}, we rely on the following result of \cite{s89}.

\begin{theorem}[{\cite{s89}}]\label{thm:fat1}
    Let $X \subseteq \position{m}$ be a set of $n$ positive integers and let $k$ be a positive integer with \[
        n > \frac{m}{k} + 1.
    \]
    Then there is an integer $1 \le h < 118k$ such that the set $X^{h}$
    contains an arithmetic progression $\mathcal{P}$ of length~$m$. 
\end{theorem}

We will need a slight adaptation of the above theorem.

\begin{lemma}[Variant of Theorem~\ref{thm:fat1}] \label{thm:varsarkozy}
  Let $X \subseteq \position{m}$ be a set of $n$ positive integers.
    Then the set $X^{\le 354 m/n} = \bigcup_{j=1}^{\lfloor 354 m/n \rfloor} X^j$
    contains an arithmetic progression $\mathcal{P}$ of length~$m$.
\end{lemma}
\begin{proof}
  Recall that we assume all our sets to be non-empty.
  If $n=1$ then we can write $X = \{x\}$. In this case, the set $X^{\le m} $ contains the arithmetic progression $\{x,2x,\ldots,mx\}$ of length $m$.
  
  If $n \ge 2$, then we set $k := \lfloor m/(n-1) \rfloor + 1$. Observe that $2 \le n \le m$ implies $k \le m/(n-1)+1 \le 3m/n$.
  Moreover, note that we have $k > m/(n-1)$ or, equivalently, $n > m/k+1$. Therefore, Theorem~\ref{thm:fat1} is applicable for $k$ and shows that the set $X^{\le 118k}$ contains an arithmetic progression of length~$m$. Finally, note that $X^{\le 118k} \subseteq X^{\le 354 m/n}$ since $k \le 3m/n$.
\end{proof}

In order to use Theorem~\ref{thm:varsarkozy}, we need to take care of two things. First, the arithmetic
progression obtained in Theorem~\ref{thm:varsarkozy} lies in $X^{\le h}$, which may use elements from the set $X$ multiple times, and thus does not correspond to subset sums.
Second, Theorem~\ref{thm:varsarkozy} assumes $X$ to be a set. (While it may seem as if both issues dissolve
for multi-sets $X$, this is the case only for multi-sets with a multiplicity of at least~$354 m/n$ {\em for every single element}.)

We tackle these two issues separately (and as in
\cite{s94}): We consider a set of integers where every element can be obtained as a sum
of two elements of $X$ in many different, disjoint ways (see Lemmas~\ref{lm:fatl2} and~\ref{lm:fatl1}). In order to obtain such a set, we first ensure that in our multi-set every number
appears equally often, that is, our multi-set has a {\em uniform} multiplicity (see Lemma~\ref{lm:im}).

\begin{definition}[Uniformity] \label{def:uniform}
  We call a multi-set $X$ \emph{uniform} if every $x \in X$ has multiplicity $\mul{X}$ in $X$.
\end{definition}

\begin{lemma} \label{lm:im}
    Let $X$ be a $\delta$-dense multi-set of size $n$.
    For any integer $0 \le r \le \log \mul{X}$, we define a subset $X_r \subseteq X$ by picking $2^r$ copies of every number with multiplicity in $[2^r,2^{r+1})$, that is, for any $x \in \mathbb{N}$ we set
    \[ \mu(x;X_r) := \begin{cases} 2^r, & \text{if } 2^r \le \mu(x;X) < 2^{r+1}, \\ 0, & \text{otherwise.} \end{cases} \] 
    There exists an integer $0 \le r \le \log \mul{X}$ such that the multi-set $X_r$ is $\delta/(4 \log^2(2\mul{X}))$-dense and has size
    \[ |X_r| \ge \frac n{2 \log(2\mul{X})}. \]
\end{lemma}
\begin{proof}
    The proof is indirect. Assume that each of the sets $X_r$
    has size $|X_r| < n / (2 \log(2\muli{X}))$. 
    By construction, at least every second element of $X$ appears in some subset $X_r$. We thus have
    \[
        \frac n2 \le \sum_{r =0}^{\lfloor \log\muli{X} \rfloor} |X_r|
        < \sum_{r =0}^{\lfloor \log\muli{X} \rfloor} \frac n{2 \log(2\muli{X})}
        \le \frac n2,
    \]
    which yields the desired contradiction.
    Hence, there exists a subset $X_r \subseteq X$ of size $|X_r| \ge n / (2 \log(2\muli{X}))$. Since $X$ is $\delta$-dense, by Observation~\ref{obs:2} we obtain that $X_r$ is $\delta / (4 \log^2(2\mul{X}))$-dense.
\end{proof}

Using Lemma~\ref{lm:im}, at the cost of some $\log(\mul{X})$-factors, we may assume that the given multi-set $X$ is {\em uniform} in the sense of \cref{def:uniform}.

We next turn to the sumset $X+X$ of a uniform multi-set $X$.

\begin{definition}[Number of Representations]
  For a \emph{set} $S$ and an integer $z$, we define $f_S(z)$ as
  the \emph{number of representations} of $z$ as the sum of two numbers in $S$, that is,
  \[ f_S(z) := |\{(x,x') \in S \times S \mid x+x'=z\}|. \]
  For a \emph{uniform multi-set} $X$ and an integer $z$, we extend this notation by defining
  \[ f_X(z) := \mul{X} \cdot f_{\supp{X}}(z).\lipicsEnd \]
\end{definition}

We start by proving basic properties of the function $f_X$.

\begin{lemma}\label{lm:fbp}
    For any uniform multi-set $X$ of size $n$, the function $f_X$ satisfies all of the following:
    \begin{enumerate}
        \item For any integer $z$ we have $f_X(z) \le n$,
        \item For any integer $z > 2 \mx{X}$ we have $f_X(z) = 0$,
        \item The sum of all values of $f_X$ is $\sum_z f_X(z) = n^2/\mul{X}$.
        \item Any integer $z$ can be written in at least $\lfloor f_X(z)/2 \rfloor$ disjoint ways as the sum of two elements of $X$, that is, there exist distinct\footnote{Here again \emph{distinct} means that any $x \in X$ may appear up to $\mu(x;X)$ times in this sequence.} elements $x_1,x'_1,\ldots,x_k,x'_k \in X$ with $x_i+x'_i = z$ for all $i$ and $k \ge \lfloor f_X(z)/2 \rfloor$. 
    \end{enumerate}
\end{lemma}
\begin{proof}
    (1) Let $S := \supp{X}$. In the definition of $f_S(z)$, after choosing $x \in X$ we must set $x' = z-x$. Thus, there are only $|S|$ options to choose from, resulting in the inequality $f_S(z) \le |S|$. For the uniform multi-set~$X$, we thus obtain $f_X(z) = \mul{X} \cdot f_S(z) \le \mul{X} \cdot |S| = |X|$.
    
    (2) Follows from the fact that the sum of two elements of $X$ is at most $2\mxi{X}$.
    
    (3) Let $S := \supp{X}$. Note that every pair $x,x' \in S$ contributes to exactly one function value of $f_S$, namely $f_S(x+x')$. Hence, we have $\sum_z f_S(z) = |S|^2$. Since $X$ is uniform, we have $|S| = n / \mul{X}$. Therefore, 
    \[ \sum_z f_X(z) = \sum_z \mul{X} f_S(z) = \mul{X} |S|^2 = n^2 / \mul{X}. \]
    
    (4) Set $S := \supp{X}$ and let $x,x' \in S$ with $x+x'=z$. First consider the case $x \ne x'$. In this case, the pairs $(x,x')$ and $(x',x)$ contribute 2 to the value $f_S(z)$, so they contribute $2 \mul{X}$ to the value $f_X(z)$. Note that we can form $\mul{X}$ many disjoint pairs between the $\mul{X}$ copies of $x$ in $X$ and the $\mul{X}$ copies of $x'$ in $X$. That is, the number of constructed pairs is half of the contribution to $f_X(z)$. 
    
    Let us turn to the case $x=x'$, that is, $x = z/2$. In this case, the pair $(x,x')$ contributes 1 to the value $f_S(z)$, so it contributes $\mul{X}$ to the value $f_X(z)$. Note that we can form $\lfloor \mul{X}/2 \rfloor$ many disjoint pairs among the $\mul{X}$ copies of $x$ in $X$. Again this is half of the contribution to $f_X(z)$, but now rounded down. In total, we have constructed $\lfloor f_X(z)/2 \rfloor$ disjoint pairs summing to $z$.
\end{proof}

\newcommand{\cB}{\mathcal{B}_{v,X}}

Next we consider buckets of numbers with an almost uniform number of representations. Informally, bucket $\cB$ contains all integers $z$ that can be written in $\Omega(v)$ many disjoint ways as the sum of two elements of $X$.

\begin{definition}[Buckets]
  For any uniform multi-set $X$ and any integer $v \ge 1$, we define the bucket
\[ \cB := \{ z \in \mathbb{N} \mid f_X(z) \ge v \}.\lipicsEnd \]
\end{definition}
Our goal is to apply Theorem~\ref{thm:varsarkozy} to an appropriate bucket $\cB$.
We list some simple observations.
\begin{lemma}\label{lm:bbp}
    Let $X$ be a uniform multi-set and let $v \ge 1$ be an integer. Then we have
    \begin{enumerate}
        \item  $|\mathcal{B}_{v,X}\!| \le 2\mxi{X}$.
        \item  If $v > |X|$ then the set $\mathcal{B}_{v,X}$ is empty.
    \end{enumerate}
\end{lemma}
\begin{proof}
    Both properties follow immediately from Lemma~\ref{lm:fbp}. For the first we use that the function $f_X$ is zero on all integers larger than $2\mxi{X}$. For the second we use that $f_X$ is bounded from
    above by $|X|$.
\end{proof}

The buckets $\cB$ can, in a limited way, remedy the multiple use
of the same element in a sum. While in general for a multi-set $X$ the set $X^h$ may contain numbers that are no subset sums of $X$, we show that for sufficiently small $h$ all numbers in $\cB^h$ also correspond to subset sums of $X$. In the proof of this statement, we use the large number of representations guaranteed by the definition of $\cB$ to avoid any multiple use of the same number.

\begin{lemma}[Compare {\cite[Lemma 2]{s94}}]\label{lm:fatl2}
    Let $X$ be a uniform multi-set.
    For any integers $v \ge 1$ and $1 \le h \le (v-1)/4$, every number in the set
    $\mathcal{B}_{v,X}^h$ can be represented as a sum of $2h$
    distinct
    elements of $X$.
\end{lemma}
\begin{proof}
    The proof is by induction on $h$. We define $\cB^0 := \{0\}$ to make the base case $h=0$ trivial. 
    
    For $h \ge 1$, consider any $w \in \cB^h$ and write $w = z+w'$ for some $z \in \cB$ and $w' \in \cB^{h-1}$.
    By the induction hypothesis, we can represent $w'$ as a sum $x_1+\ldots+x_{2h-2}$ of distinct elements of $X$.
    By the definition of $\cB$ we have $f_X(z) \ge v$, so by Lemma~\ref{lm:fbp}.4 we can find $\lfloor v/2 \rfloor \ge (v-1)/2 \ge 2h$ disjoint representations of the number $z$ as the sum of two numbers in $X$. By the pigeonhole principle, at least one of these $2h$ many representation of $z$ does not contain any of the $2h-2$ many numbers $x_1,\ldots,x_{2h-2}$. We pick such a representation $z = x_{2h-1}+x_{2h}$ to obtain a representation of $w$ as a sum $x_1+\ldots+x_{2h}$ of $2h$ distinct elements of $X$.
\end{proof}

The above lemma shows that plugging the set $\cB$ into Theorem~\ref{thm:varsarkozy} yields knowledge about the subset sums of $X$, despite the formulation of Theorem~\ref{thm:varsarkozy} allowing to pick summands multiple times. 
In order to use Theorem~\ref{thm:varsarkozy} effectively, we need to pick a bucket $\cB$ of large size. We next prove existence of such a bucket.

\begin{lemma}[Compare {\cite[Lemma 1]{s94}}]\label{lm:fatl1}
    Let $X$ be a uniform multi-set of size $n$ that is 7-dense.
    Then there is an integer $1 < v \le n$ such that the set
    $\mathcal{B}_{v,X}$ satisfies \[
        |\mathcal{B}_{v,X}| \ge \frac{n}{3 \muli{X}} +
    \frac{n^2}{3 \,v\, \muli{X}\log(2n)}.\]
\end{lemma}
\begin{proof}
    The proof is indirect. Assume that for every integer $1 < v \le n$ we have \[
        |\mathcal{B}_{v,X}| < \frac{n}{3 \muli{X}} +
    \frac{n^2}{3 \,v\, \muli{X}\log(2n)}.
    \]
    From the construction of the buckets $\cB$, we observe the following identity:
    \[ \sum_z f_X(z) = \sum_{v \ge 1} v \cdot \big(|\cB| - |\mathcal{B}_{v+1,X}|\big). \]
    By telescoping this sum and by using $\sum_z f_X(z) = n^2 / \mul{X}$ from Lemma~\ref{lm:fbp}.3, we arrive at
    \[ \frac {n^2}{\mul{X}} = \sum_{v \ge 1} |\cB|. \]
    We bound the right hand side by using $|\cB| = 0$ for $v >n$ (by Lemma~\ref{lm:bbp}.2), the assumed upper bound for $1 < v \le n$, and $|\cB| \le 2\mx{X}$ for $v=1$ (by Lemma~\ref{lm:bbp}.1). 
    This yields
    \begin{align*} 
      \frac {n^2}{\mul{X}} &\le 2\mx{X} + \sum_{v = 2}^{n} \bigg( \frac{n}{3 \muli{X}} +
    \frac{n^2}{3 \,v\, \muli{X}\log(2n)} \bigg) \\
      &\le 2\mx{X} + \frac{n^2}{3 \mul{X}} + \frac{n^2}{3 \mul{X} \log(2n)} \cdot \sum_{v=1}^n \frac 1v.
    \end{align*}
    We now use the standard fact $\sum_{v=1}^n 1/v \le 1 + \log(n) = \log(2n)$ to obtain
    \[ \frac {n^2}{\mul{X}} \le 2\mx{X} + \frac{2 n^2}{3 \mul{X}}. \]
    Finally, we use that the multi-set $X$ is 7-dense, so that $n^2 \ge 7 \mul{X} \mx{X}$, which yields the desired contradiction
    \[ \frac {n^2}{\mul{X}} \le \Big(\frac 27 + \frac 23 \Big) \cdot \frac {n^2}{\mul{X}}. \qedhere \]
\end{proof}

Combining Lemmas \ref{thm:varsarkozy}, \ref{lm:im}, \ref{lm:fatl2}, and \ref{lm:fatl1}
we now prove the main theorem of this section.

\tmap*

\begin{proof}
  In order for the theorem statement to be non-trivial we must have
  \[ 2\mx{X} \le \frac{n^2}{33984 \mul{X} \log (2n) \log^2(2\mul{X})}. \]
  Rearranging this shows that $X$ must be $\delta$-dense for 
  \[ \delta := 67968 \log(2n) \log^2(2\mul{X}). \]  
  We first apply Lemma~\ref{lm:im} to obtain a subset $X' := X_r \subseteq X$ such that $X'$ is a uniform multi-set of size~$n'$ that is $\delta'$-dense, where
  \begin{align*}
    n \ge n' \ge \frac n{2\log(2\mul{X})}, \qquad
    \delta' = \frac {\delta}{4 \log^2(2\mul{X})} \ge 7.
  \end{align*}
  Next, we apply Lemma~\ref{lm:fatl1} to obtain an integer $1 < v \le n'$ such that
  \[ |\mathcal{B}_{v,X'}| \ge \frac{n'}{3\muli{X}} +
    \frac{n'^2}{3\, v\, \muli{X}\log(2n)}.\]
  By Lemma~\ref{lm:fbp}.2, for any $m \ge 2 \mx{X}$ we have $\mathcal{B}_{v,X'} \subseteq \position{m}$. Thus, Theorem~\ref{thm:varsarkozy} yields that the set $\mathcal{B}_{v,X'}^{\le h}$ contains an arithmetic progression $\mathcal{P}$ of length $m$, for $h := 354m/|\mathcal{B}_{v,X'}|$. Using our bounds on $|\mathcal{B}_{v,X'}|$ and $n'$, we obtain
  \[ h = \frac{354m}{|\mathcal{B}_{v,X'}|} \le 354m \cdot \frac{3 \,v\, \muli{X}\log(2n)}{n'^2} \le 354m \cdot \frac{12 \,v\, \muli{X}\log(2n) \log^2(2\mul{X})}{n^2} \le \frac v8, \]
  where the last step uses the assumption on $m$.
  Since $v > 1$ is an integer, we have $v-1 \ge v/2$, so we can further bound
  \[ h \le \frac{v-1}4. \]
  Therefore, Lemma~\ref{lm:fatl2} is applicable and implies that the arithmetic progression $\mathcal{P}$ also appears as a subset of $\ssm{X}$. 
  Moreover, Lemma~\ref{lm:fatl2} shows that every element of $\mathcal{P}$ can be written as the sum of at most $2h$ distinct elements of $X$. We now bound differently from before:
  \[ h = \frac{354m}{|\mathcal{B}_{v,X'}|} \le 354m \cdot \frac {3 \mul{X}}{n'} \le 354m \cdot \frac{6 \mul{X} \log(2\mul{X})}{n}. \]
  This shows that every element of $\mathcal{P}$ can be obtained as the sum of at most $2h \le 4248 m\, \mul{X} \log(2\mul{X})/n$ distinct elements of $X$.
  In particular, we obtain
  \begin{align} \label{eq:ojgohsdjfkasf}
    \mx{\mathcal{P}} \le 2h \cdot \mx{X} \le 4248 m\, \mul{X} \mx{X} \log(2\mul{X})/n.
  \end{align}
  
  Denote by $s$ the step size of the arithmetic progression $\mathcal{P}$. Then we have $m \cdot s \le \mx{\mathcal{P}}$ (here we use that $\mathcal{P}$ is of the form $\{a+s,a+2s,\ldots,a+m\cdot s\}$). 
  Together with inequality (\ref{eq:ojgohsdjfkasf}), this yields
  \[ s \le 4248 \mul{X} \mx{X} \log(2\mul{X})/n. \qedhere \]
\end{proof}

\subsubsection{Constructing a Decomposition} \label{sec:decomposition}

We use the tools from the last two sections to decompose $X$ as follows.

\begin{theorem} \label{thm:decomposition}
  Let $X$ be a $\delta$-dense multi-set of size $n$ that has no $\alpha$-almost divisor, where 
  \begin{align*}
    \delta &= 1699200 \log (2n) \log^2(2\mul{X}), \\
    \alpha &= 42480 \log(2\mul{X}).
  \end{align*}
  There exists a partitioning $X = R \cup A \cup G$ and an integer $s \le 42480 \cdot \mul{X} \sm{X} \log(2\mul{X}) / n^2$ such that
  \begin{itemize}
    \item the set $\ssm{R}$ is $s$-complete, that is, $\ssm{R} \bmod s = \mathbb{Z}_s$,
    \item the set $\ssm{A}$ contains an arithmetic progression $\mathcal{P}$ of length $2 \mx{X}$ and step size $s$ satisfying $\mx{\mathcal{P}} \le 84960 \mul{X} \mx{X} \sm{X} \log(2\mul{X}) / n^2$,
    \item the multi-set $G$ has sum $\sm{G} \ge \sm{X}/2$.
  \end{itemize}   
\end{theorem}
\begin{proof}
Given the multi-set $X$, we first use Lemma~\ref{lm:t34} to obtain the subset $R \subseteq X$. From the remaining elements $X \setminus R$ we pick the smallest $\lfloor n/4 \rfloor$ elements to form the set~$A$. 
We call the remaining set $G := (X \setminus R) \setminus A$. This yields the partitioning $X = R \cup A \cup G$ (in the sense of $\mu(x;X) = \mu(x;R)+\mu(x;A)+\mu(x;G)$). 

\medskip
In the following we analyze the properties of this decomposition.
\begin{claim} \label{cla:propR}
The multi-set $R$ satisfies:
\begin{itemize}
  \item $|R| \le n/4$,
  \item $\sm{R} \le \sm{X}/4$, and
  \item $R$ is $d$-complete for any $d \le \alpha \cdot \mul{X} \sm{X} / n^2 = 42480 \cdot \mul{X} \sm{X} \log(2\mul{X}) / n^2$.
\end{itemize} 
\end{claim}
\begin{claimproof}
  The first two properties follow directly from Lemma~\ref{lm:t34} and the inequality $\delta \ge 32 \alpha \log(2n)$, which follows from the definitions of $\delta$ and $\alpha$.
  
  Lemma~\ref{lm:t34} also shows that for any $1 < d \le \alpha \cdot \mul{X} \sm{X} / n^2$ we have $|\overline{R(d)}| \ge d$. The third claim now follows from Lemma~\ref{lm:naco}.
\end{claimproof}

\begin{claim} \label{cla:propA}
The multi-set $A$ satisfies:
\begin{itemize}
  \item $n/5 \le |A| \le n/4$,
  \item $\mx{A} \le 2 \sm{X}/n$,
  \item $\ssm{A}$ contains an arithmetic progression $\mathcal{P}$ of length $2 \mx{X}$ and step size 
  \[ s \le 42480 \cdot \frac{\mul{X} \sm{X}}{n^2} \log(2\mul{X}). \]
  The arithmetic progression $\mathcal{P}$ moreover satisfies
  \[ \mx{\mathcal{P}} \le 84960 \cdot \frac{\mul{X} \mx{X} \sm{X}}{n^2} \log(2\mul{X}). \]
\end{itemize} 
\end{claim}
\begin{claimproof}
  By definition of $\delta$-density we obtain $n^2 \ge \delta \ge 225$ and thus $n \ge 15$. This implies $|A| = \lfloor n/4 \rfloor \ge (n-3)/4 \ge n/5$, which proves the first claim.
  
  Since $R$ picks at most $n/4$ elements from $X$ and $A$ picks the $\lfloor n/4 \rfloor$ many smallest remaining elements, it follows that every elements in $A$ is bounded from above by the median of $X$. Since $X$ contains at least $n/2$ elements that are larger than or equal to the median, the median is bounded from above by $\sm{X}/(n/2) = 2\sm{X}/n$. Hence, we have $\mx{A} \le 2 \sm{X}/n$.
  
  For the last claim, we apply Theorem~\ref{tm:ap} to the multi-set $A$ and $m := 2 \mx{X}$. Let us check the preconditions of this theorem. We clearly have $m = 2\mx{X} \ge 2\mx{A}$. Moreover, we have
  \[ \frac{|A|^2}{33984 \mul{A} \log (2|A|) \log^2(2\mul{A})} \ge \frac{n^2}{5^2 \cdot 33984 \mul{X} \log (2n) \log^2(2\mul{X})} \ge 2 \mx{X} = m, \]
  where we used the assumption that $X$ is $\delta$-dense for $\delta \ge 1699200 \log (2n) \log^2(2\mul{X})$.
  Thus, Theorem~\ref{tm:ap} is applicable to $(A,m)$ and yields an arithmetic progression $\mathcal{P}$ in $\ssm{A}$ of length $m$ and step size 
  \[ s \le 4248 \cdot \frac{\mul{A} \mx{A}}{|A|} \log(2\mul{A}) \le 5 \cdot2 \cdot 4248 \cdot \frac{\mul{X} \sm{X}}{n^2} \log(2\mul{X}) = 42480 \cdot \frac{\mul{X} \sm{X}}{n^2} \log(2\mul{X}), \]
  where we used the properties $|A| \ge n/5$ and $\mx{A} \le 2\sm{X}/n$.
  Moreover, from Theorem~\ref{tm:ap} we also obtain 
  \[ \mx{\mathcal{P}} \le 4248 \cdot \frac{m \mul{A} \mx{A}}{|A|} \log(2\mul{A}) \le 5\cdot 2 \cdot 4248 \cdot 2 \cdot \frac{\mx{X} \mul{X} \sm{X}}{n^2} \log(2\mul{X}) = 84960 \cdot \frac{\mul{X} \mx{X} \sm{X}}{n^2} \log(2\mul{X}). \]
\end{claimproof}

\begin{claim} \label{cla:propG}
The multi-set $G$ satisfies $\sm{G} \ge \sm{X}/2$.
\end{claim}
\begin{claimproof}
  Since $A$ picks the $\lfloor n/4 \rfloor$ smallest elements of $X\setminus R$, and since $|X\setminus R| \ge \frac 34 n$ by Claim~\ref{cla:propR}, we have
  \[ \sm{A} \le \frac{|A|}{|X\setminus R|} \cdot \sm{X\setminus R} \le \frac 13 \sm{X\setminus R} = \frac 13 (\sm{X} - \sm{R}). \]
  Using $\sm{R} \le \sm{X}/4$ from Claim~\ref{cla:propR}, we obtain
  \[ \sm{A} + \sm{R} \le \frac 13 (\sm{X} - \sm{R}) + \sm{R} = \frac 13 \sm{X} + \frac 23 \sm{R} \le \Big(\frac 13 + \frac 16\Big) \sm{X} = \frac{\sm{X}}2. \]
  Therefore, $\sm{G} = \sm{X} - \sm{R} - \sm{A} \ge \sm{X}/2$.
\end{claimproof}

Note that since the multi-set $R$ is $d$-complete for each small $d$, in particular $R$ is also $s$-complete. Hence, Claims~\ref{cla:propR}, \ref{cla:propA}, and \ref{cla:propG} finish the proof of Theorem~\ref{thm:decomposition}.
\end{proof}

\subsubsection{Proof of the Structual Part} \label{sec:struct_proof}

Finally, we are ready to prove the structural part.

\thmmainstructformal*

\begin{proof}
  We want to show that any target number $t \in \fragment{\lambda_X}{\sm{X}-\lambda_X}$ is also a subset sum of $X$. By symmetry, it suffices to prove the claim for $t \le \sm{X}/2$. 
  
  We construct the partitioning $X = R \cup A \cup G$ from Theorem~\ref{thm:decomposition}. 
  We denote the arithmetic progression $\mathcal{P} \subseteq \ssm{A}$ by $\mathcal{P} = \{a+s,a+2s,\ldots,a+2\mx{X} s\}$.
  
  We construct a subset summing to $t$ as follows.
  First, we pick a subset $G' \subseteq G$ by greedily adding elements until 
  \[ t - a - s\cdot (\mx{X}+1) - \mx{X} < \sm{G'} \le t - a - s\cdot (\mx{X}+1). \]
  This is possible because this range for $\sm{G'}$ has length $\mx{X}$, and because we have $t \le \sm{X}/2 \le \sm{G}$ and 
  \[ t \;\ge\; \lambda_X \;\ge\; (84960 + 2\cdot 42480) \cdot \mul{X} \mx{X} \sm{X} \log(2\mul{X}) / n^2 \;\ge\; \mx{\mathcal{P}} + 2 s\, \mx{X} \;\ge\; a + s\cdot(\mx{X}+1). \]
  
  Next we pick a subset $R' \subseteq R$ that sums to $(t - \sm{G'} - a)$ modulo $s$. This is possible because $R$ is $s$-complete. We can assume that $R'$ has size $|R'| \le s$, since otherwise some subset of $R'$ sums to 0 modulo~$s$ and can be removed (the details of this argument were explained in the proof of Claim~\ref{cla:wlogY}).
  In particular, we can assume $\sm{R'} \le s \cdot \mx{X}$. 
  
  We thus have $t - \sm{G' \cup R'} \equiv a \pmod s$ and 
  \[t - a - s\cdot (\mx{X}+1) - \mx{X} < \sm{G' \cup R'} \le t - a - s, \]
  or, equivalently,
  \[ a + s \le t - \sm{G' \cup R'} < a + s\cdot (\mx{X}+1) + \mx{X}. \]
  Note that for any positive integers $x,y$ we have $(x-1)(y-1) \ge 0$. Rearranging this yields $x+y \le xy+1$, or equivalently $x(y+1) + y \le 2xy+1$. Using this, we obtain
  \[ a + s \le t - \sm{G' \cup R'} \le a + 2 \mx{X} s. \]
  Note that this is exactly the range of the elements of the arithmetic progression $\mathcal{P}$. Moreover, since the remaining target $t - \sm{G' \cup R'}$ is of the form $a + k s$ for some integer $k$, we can pick a subset $A' \subseteq A$ that gives the appropriate element of the arithmetic progression $\mathcal{P}$ and thus yields the desired $t = \sm{G' \cup R'\cup A'}$. 
\end{proof}

This finishes the proof of the structural part, and thus of the algorithm.

\section{Fine-Grained Lower Bound} \label{sec:lowerbound}
\tikzset{numbox/.style={draw=black!50, text width=2cm, align=right, text height=.3cm, inner sep=2.3pt}}
\tikzset{numboxs/.style={draw=black!50, text width=1cm, align=right, text height=.3cm, inner sep=2.3pt}}
\tikzset{numboxl/.style={draw=black!50, text width=4cm, align=right, text height=.3cm, inner sep=2.3pt}}
\tikzset{numspc/.style={text width=2cm, align=center, text height=.3cm}}
\tikzset{numspcs/.style={text width=1cm, align=center, text height=.3cm}}

Before presenting our conditional lower bound, we introduce and discuss our hardness assumptions.

\subsection{Hardness Assumptions}

\subsubsection{Strong Exponential Time Hypothesis}

We first consider the classic Satisfiability problem, more precisely the \kSAT problem.

\begin{problem}[\kSAT]
    Given a $k$-CNF formula $\varphi$ on $N$ variables and $M$ clauses, decide whether
    $\varphi$ is satisfiable, that is, decide whether there is an assignment of {\tt true}
    or {\tt false} to the variables such that $\varphi$ is satisfied.\lipicsEnd
\end{problem}
The Strong Exponential Time Hypothesis was introduced by Impagliazzo, Paturi, and Zane and essentially postulates that there is no exponential improvement over exhaustive search for the \kSAT problem. This is the most widely used hardness assumption in fine-grained complexity theory~\cite{williams2018some}.
\begin{conjecture}[Strong Exponential Time Hypothesis (SETH) \cite{ip01,cip09}]
    For any $\varepsilon > 0$ there is an integer $k \ge 3$ such that
    \kSAT cannot be solved in time $O(2^{(1-\varepsilon)N})$.\lipicsEnd
\end{conjecture}

\subsubsection{Strong k-Sum Hypothesis}
\label{sec:lb:hypo:ksum}

\begin{problem}[\kSum]
  Given a set $Z \subseteq \position{U}$ of $N$ integers and a target $T$, decide whether there exist $z_1,\ldots,z_k \in Z$ with $z_1+\ldots+z_k = T$.\lipicsEnd
\end{problem}

The \kSum problem has classic algorithms running in time $\Oh(N^{\lceil k/2 \rceil})$ (via meet-in-the-middle) and in time $\tOh(U)$ (via Fast Fourier transform). 
The (standard) $k$-Sum Hypothesis postulates that the former algorithm cannot be improved by polynomial factors, i.e., \kSUM has no 
$\Oh(N^{\lceil k/2 \rceil-\eps})$-time algorithm for any $\eps > 0$~\cite{GajentaanO95}. 
Note that both algorithmic approaches yield the same running time when $U \approx N^{\lceil k/2 \rceil}$. The Strong $k$-Sum Hypothesis postulates that even in this special case both algorithms cannot be improved by poynomial factors.

\begin{conjecture}[Strong k-Sum Hypothesis~\cite{AmirCLL14,AbboudBBK17}]
  For any $k \ge 3$ and $\eps > 0$, the \kSum problem restricted to $U = N^{\lceil k/2 \rceil}$ cannot be solved in time $\Oh(N^{\lceil k/2 \rceil-\eps})$.\lipicsEnd
\end{conjecture}

\subsubsection{Intermediate Hypothesis}

In this paper, we introduce and make use of the following hypothesis.

\begin{conjecture}[Intermediate Hypothesis]
  For any constants $\alpha,\eps > 0$ there exists a constant $k \ge 3$ such that \kSUM restricted to $N \le U^\alpha$ cannot be solved in time $\Oh(U^{1-\eps})$.\lipicsEnd
\end{conjecture}

We call this hypothesis ``intermediate'' because it not as strong as the Strong k-Sum Hypothesis. Indeed, the latter implies the former.

\begin{lemma} \label{lem:3sum_intermediate}
  The Strong k-Sum Hypothesis implies the Intermediate Hypothesis.
\end{lemma}
\begin{proof}
  We show that if the Intermediate Hypothesis fails then the Strong k-Sum Hypothesis fails.
  
  If the Intermediate Hypothesis fails, then there exist $\alpha,\eps > 0$ such that for all $k \ge 3$ the \kSUM problem restricted to $N \le U^\alpha$ can be solved in time $\Oh(U^{1-\eps})$. In particular, this holds for $k := \lceil 2/\alpha \rceil$. For this value of $k$, we have $U^\alpha \ge U^{2/k} \ge U^{1/\lceil k/2 \rceil}$. Hence, any \kSUM instance with $U = N^{\lceil k/2 \rceil}$ satisfies $N \le U^\alpha$.
  In particular, using the assumed algorithm, \kSUM restricted to $U = N^{\lceil k/2 \rceil}$ can be solved in time $\Oh(U^{1-\eps}) = \Oh(N^{\lceil k/2 \rceil - \eps'})$ for $\eps' := \eps \cdot \lceil k/2 \rceil > 0$, so the Strong k-SUM Hypothesis fails.
\end{proof}

Moreover, the Intermediate Hypothesis also follows from the Strong Exponential Time Hypothesis.

\begin{lemma} \label{lem:seth_intermediate}
  SETH implies the Intermediate Hypothesis.
\end{lemma}

This follows from the following theorem by Abboud et al.~\cite{AbboudBHS19}.
\begin{theorem}[\cite{AbboudBHS19}] \label{thm:abhs19}
  Assuming SETH, for any $\eps > 0$ there exists $\delta > 0$ such that for any $k$ the \kSUM problem is not in time $\Oh(T^{1-\eps} N^{\delta k})$. \lipicsEnd
\end{theorem}

\begin{proof}[Proof of Lemma~\ref{lem:seth_intermediate}]
  We show that if the Intermediate Hypothesis fails then SETH fails.
  
  If the Intermediate Hypothesis fails, then there exist $\alpha,\eps >0$ such that for any $k \ge 3$ we can solve \kSUM restricted to $N \le U^\alpha$ in time $\Oh(U^{1-\eps})$. 
  We claim that, without any restriction on $N$, we can then solve \kSUM in time $\Oh(U^{1-\eps} + N^{1/\alpha} \textup{polylog} N)$.
  Indeed, for $N \le U^\alpha$ we assumed running time $\Oh(U^{1-\eps})$, and for $N > U^\alpha$ using the standard algorithm based on Fast Fourier Transform we solve \kSUM in time $\tOh(U) = \tOh(N^{1/\alpha})$. 
  We roughly bound this time by $\Oh(U^{1-\eps} + N^{1/\alpha} \textup{polylog} N) = \Oh(U^{1-\eps} N^{2/\alpha})$. 
  
  Note that for \kSUM we can assume without loss of generality that $U \le T$, since input numbers larger than $T$ can be ignored. 
  Therefore, we can bound the running time of our \kSUM algorithm by $\Oh(U^{1-\eps} N^{2/\alpha}) = \Oh(T^{1-\eps} N^{2/\alpha})$. 
  
  Now it is useful to consider the contraposition of Theorem~\ref{thm:abhs19}: 
  If there exists $\eps > 0$ such that for all $\delta > 0$ there exists $k$ such that \kSUM is in time $\Oh(T^{1-\eps} N^{\delta k})$, then SETH fails. 
  We note that our $\Oh(T^{1-\eps} N^{2/\alpha})$-time algorithm for \kSUM satisfies the precondition of this statement, by picking the same value for $\eps$ and setting $k := \lceil 2/(\alpha \delta) \rceil$. 
  Hence, we showed that SETH fails.
\end{proof}

By the above two lemmas, if we want to prove a lower bound based on the Strong Exponential Time Hypothesis and the Strong k-Sum Hypothesis, then it suffices to prove a lower bound based on the Intermediate Hypothesis.

\subsection{A Lower Bound for Subset Sum}
\label{sec:lb:lb}

We can now present our lower bound. Throughout this section we only consider the set case, so we let $X$ be a set of size $n$, in particular $\mul{X}=1$. Recall that in this case \SubsetSum can be solved in time $\tOh(n)$ if $t \gg \mx{X} \sm{X} / n^2$ (Theorem~\ref{thm:intro_main_algo}). 

Our goal is to show that this regime essentially characterizes all near-linear-time settings, that is, dense \SubsetSum is not in near-linear time for $t \ll \mx{X} \sm{X} / n^2$. 
To show that we do not miss any setting, we consider a notion of \emph{parameter settings} similarly as in~\cite{BringmannK18}: For the parameters $t,\mx{X},\sm{X}$ we fix corresponding exponents $\tau,\xi,\sigma$, and we focus on \SubsetSum instances $(X,t)$ that satisfy $t = \Theta(n^\tau)$, $\mx{X} = \Theta(n^\xi)$, and $\sm{X} = \Theta(n^\sigma)$. This defines a \emph{slice} or \emph{parameter setting} of the \SubsetSum problem. Our goal is to prove a conditional lower bound for each parameter setting in which the near-linear time algorithms do not apply.

We note that some choices of the exponents $\tau,\xi,\sigma$ are contradictory, in the sense that there exist no (or only finitely many) instances satisfying $t = \Theta(n^\tau)$, $\mx{X} = \Theta(n^\xi)$, and $\sm{X} = \Theta(n^\sigma)$. 
Additionally, some assumptions can be made that hold without loss of generality. 
Specifically, we call any parameter setting $(\tau,\xi,\sigma)$ \emph{non-trivial} if it satisfies all of the following justified inequalities:
\begin{itemize}
  \item $\xi \ge 1$: For any set $X$ of $n$ positive integers we have $\mx{X} \ge n$.
  \item $\sigma \ge 2$: For any set $X$ of $n$ positive integers we have $\sm{X} \ge \sum_{i=1}^n i = \Omega(n^2)$.
  \item $\sigma \le 1 + \xi$: Any set $X$ satisfies $\sm{X} \le n \cdot \mx{X}$.
  \item $\tau \ge \xi$: Since any numbers in $X$ larger than $t$ can be ignored, we can assume $t \ge \mx{X}$.
  \item $\tau \le \sigma$: If $t > \sm{X}$ then there is no solution, so the problem is trivial.
\end{itemize}
Recall that we only want to prove a super-linear lower bound in the regime $t \ll \mx{X} \sm{X} / n^2$. We call a parameter setting \emph{hard} if it satisfies the corresponding inequality on the exponents:
\begin{itemize}
  \item $\tau < \xi + \sigma - 2$. 
\end{itemize}
Our goal is to show a lower bound of the form $n^{1+\Omega(1)}$ for each hard non-trivial parameter setting.

This discussion is summarized and formalized by the following definition.
\begin{definition}
  A {\em parameter setting} is a tuple $(\tau,\xi,\sigma) \in \mathbb{R}^3$. A parameter setting is called {\em non-trivial} if $\sigma \ge 2$ and $1 \le \xi \le \tau \le \sigma \le 1 + \xi$. A parameter setting is called {\em hard} if $\tau < \xi + \sigma - 2$. 
  
  For a parameter setting $(\tau,\xi,\sigma)$ and a constant $\rho \ge 1$ we define \SubsetSumN$^\rho(\tau,\xi,\sigma)$ as the set of all \SubsetSum instances $(X,t)$ for which the quotients $t / |X|^\tau$, $\mx{X} / |X|^\xi$, and $\sm{X} / |X|^\sigma$ all lie in the interval $[1/\rho, \rho]$.
  In some statements we simply write \SubsetSumN$(\tau,\xi,\sigma)$ to abbreviate that there exists a constant $\rho \ge 1$ such that the statement holds for \SubsetSumN$^\rho(\tau,\xi,\sigma)$.\lipicsEnd
\end{definition}

Note that we can express the running time of an algorithm solving \SubsetSumN$(\tau,\xi,\sigma)$ either in terms of $n,t,\mx{X},\sm{X}$ or in terms of $n,n^\tau,n^\xi,n^\sigma$, both views are equivalent.
Our main result of this section is:

\begin{theorem}[Lower Bound with Parameter Settings] \label{thm:seclb_main}
  Assuming the Intermediate Hypothesis,
  for any non-trivial parameter setting $(\tau,\xi,\sigma)$ there is a constant $\rho \ge 1$ such that for any $\eps >0$ the problem \SubsetSumN$^\rho(\tau,\xi,\sigma)$ cannot be solved in time $\Oh\big((\mx{X} \sm{X}/(n t))^{1-\eps}\big) = \Oh\big(n^{(\xi+\sigma-\tau-1)(1-\eps)}\big)$.
  \lipicsEnd
\end{theorem}

By Lemmas~\ref{lem:3sum_intermediate} and \ref{lem:seth_intermediate}, the same lower bound also holds under the Strong Exponential Time Hypothesis and under the Strong k-Sum Hypothesis.
Ignoring the notion of parameter settings, we have thus shown that \SubsetSum cannot be solved in time $\Oh\big((\mx{X} \sm{X}/(n t))^{1-\eps}\big)$ for any $\eps > 0$, unless the Strong Exponential Time Hypothesis and the Strong k-Sum Hypothesis both fail. This proves Theorem~\ref{thm:intro_lowerbound}.

\medskip
Note that Theorem~\ref{thm:seclb_main} is trivial when $n^{\xi+\sigma-\tau-1} \le n$, since it is then subsumed by the trivial lower bound of $\Omega(n)$ to read the input. 
Therefore, we only need to prove the theorem statement for \emph{hard} non-trivial parameter settings.

We prove Theorem~\ref{thm:seclb_main} by a reduction from \kSUM to any hard non-trivial parameter setting of \SubsetSum. The reduction transforms the hypothesized time complexity $U^{1 - o(1)}$ of \kSUM into a lower bound of $n^{\xi+\sigma-\tau-1-o(1)}$ for \SubsetSum. 

\begin{lemma}[The Reduction] \label{lem:thereduction}
  Let $(\tau,\xi,\sigma)$ be a hard non-trivial parameter setting and fix $k \ge 3$. Set $\alpha := 1/(\xi+\sigma-\tau-1)$.
  Given an instance $(Z,T)$ of \kSUM with $Z \subseteq \position{U}$ and $|Z| \le U^\alpha$, in time $\Oh(U^\alpha)$ we can construct an equivalent instance $(X,t)$ of \SubsetSumN$(\tau,\xi,\sigma)$ with $|X| = \Theta(U^{\alpha})$. \lipicsEnd
\end{lemma}

This reduction easily implies Theorem~\ref{thm:seclb_main}.
\begin{proof}[Proof of Theorem~\ref{thm:seclb_main}]
  Assume that some non-trivial parameter setting \SubsetSumN$(\tau,\xi,\sigma)$ can be solved in time $\Oh(n^{(\xi+\sigma-\tau-1)(1-\eps)})$ for some $\eps > 0$. 
  Since reading the input requires time $\Omega(n)$, we must have $\xi+\sigma-\tau-1 > 1$, which is equivalent to $\alpha := 1/(\xi+\sigma-\tau-1) < 1$. 
  Pick any $k \ge 3$ and an instance $(Z,T)$ of \kSUM with $Z \subseteq \position{U}$ and $|Z| \le U^\alpha$. Run the reduction from Lemma~\ref{lem:thereduction} to produce an equivalent \SubsetSum instance $(X,t)$. The reduction itself runs in time $\Oh(U^\alpha)$.
  Now we use the assumed algorithm to solve the instance $(X,t)$ in time $\Oh(|X|^{(\xi+\sigma-\tau-1)(1-\eps)}) = \Oh(U^{1-\eps})$. Since $(X,t)$ is equivalent to $(Z,T)$, we have thus solved the given instance $(Z,T)$ in time $\Oh(U^\alpha + U^{1-\eps}) = \Oh(U^{1-\eps'})$ for $\eps' := \min\{\eps,1-\alpha\} > 0$. This violates the Intermediate Hypothesis.
\end{proof}

It remains to design the reduction.

\subsection{The Reduction}

In this section we prove Lemma~\ref{lem:thereduction}.

\def\constt{2}
\def\const{4 k}
\def\Const{8 k^2}
\begin{figure}[t]
    \centering
    \begin{tikzpicture}
        \node[anchor=west] (a) at (-1.25, 1.5) {$\constt\,U^\beta$};
        \draw (a.west) -- ($(a.west) + (0, -.5)$);
        \node[anchor=west] (a) at (-2.75, 1.5) {$\const\, U^{\beta}$};
        \draw (a.west) -- ($(a.west) + (0, -.5)$);
        \node[anchor=west] (a) at (-5.25, 1.5) {$\Const\, U^{1 + \beta}$};
        \draw (a.west) -- ($(a.west) + (0, -.5)$);
        \node[anchor=west] (a) at (-7.75, 1.5) {$\Const\, U^{1 + \beta + \gamma}$};
        \draw (a.west) -- ($(a.west) + (0, -.5)$);

        \node[numbox] (1) at (0,.6) {\phantom{$X_1$}0};
        \node[numbox] at (0,0) {\phantom{$X_1$}0};
        \node[numspc] at (0,-.6) {$\vdots$};
        \node[numbox] (2) at (0,-1.2) {\phantom{$X_1$}0};
        \draw [decorate,decoration={brace,amplitude=4pt},yshift=0pt]
            ($(1.north east) + (.2,.04)$) -- node[right, align=left, xshift=5] {$X_1$}
            ($(2.south east) + (.2,-.05)$);

        \node[numboxs] at (-2,.6) {\phantom{$X_1$}1};
        \node[numboxs] at (-2,0)  {\phantom{$X_1$}1};
        \node[numspcs] at (-2,-.6){$\vdots$};
        \node[numboxs] at (-2,-1.2){\phantom{$X_1$}1};

        \node[numbox, align=right] at (-4,.6)  {$z_1$};
        \node[numbox, align=right] at (-4,0)   {$z_2$};
        \node[numspc] at (-4,-.6) {$\vdots$};
        \node[numbox, align=right] at (-4,-1.2){$z_{|Z|}$};

        \node[numbox] at (-6.5,.6)  {\phantom{$X_1$}1};
        \node[numbox] at (-6.5,0)   {\phantom{$X_1$}1};
        \node[numspc] at (-6.5,-.6) {$\vdots$};
        \node[numbox] at (-6.5,-1.2){\phantom{$X_1$}1};

        \draw[black!50, dashed] (-8, -1.7) -- (1.5, -1.7);

        \node[numbox] (x1) at (0,-5) {\phantom{$X_1$}0};
        \node[numbox] at (0,-5.6) {\phantom{$X_1$}0};
        \node[numspc] at (0,-6.2) {$\vdots$};
        \node[numbox] (x2) at (0,-6.8) {\phantom{$X_1$}0};
        \draw [decorate,decoration={brace,amplitude=4pt},yshift=0pt]
            ($(x1.north east) + (.2,.04)$) -- node[right, align=left, xshift=5] {$X_3$}
            ($(x2.south east) + (.2,-.05)$);

        \node[numboxs] at (-2,-5) {\phantom{$X_1$}0};
        \node[numboxs] at (-2,-5.6)  {\phantom{$X_1$}0};
        \node[numspcs] at (-2,-6.2){$\vdots$};
        \node[numboxs] at (-2,-6.8){\phantom{$X_1$}0};

        \node[numbox] at (-4,-5)  {\phantom{$X_1$}1};
        \node[numbox] at (-4,-5.6)   {\phantom{$X_1$}2};
        \node[numspc] at (-4,-6.2) {$\vdots$};
        \node[numbox] at (-4,-6.8){\phantom{$X_1$}$U^\alpha$};

        \node[numbox] at (-6.5,-5)  {\phantom{$X_1$}1};
        \node[numbox] at (-6.5,-5.6)   {\phantom{$X_1$}1};
        \node[numspc] at (-6.5,-6.2) {$\vdots$};
        \node[numbox] at (-6.5,-6.8){\phantom{$X_1$}1};

        \draw[black!50, dashed] (-8, -4.5) -- (1.5, -4.5);

        \node[numbox] (v1) at (0,-2.2) {\phantom{$X_1$}1};
        \node[numbox] at (0,-2.8) {\phantom{$X_1$}1};
        \node[numspc] at (0,-3.4) {$\vdots$};
        \node[numbox] (v2) at (0,-4) {\phantom{$X_1$}1};
        \draw [decorate,decoration={brace,amplitude=4pt},yshift=0pt]
            ($(v1.north east) + (.2,.04)$) -- node[right, align=left, xshift=5] {$X_2$}
            ($(v2.south east) + (.2,-.05)$);

        \node[numboxs] at (-2,-2.2) {\phantom{$X_1$}0};
        \node[numboxs] at (-2,-2.8)  {\phantom{$X_1$}0};
        \node[numspcs] at (-2,-3.4){$\vdots$};
        \node[numboxs] at (-2,-4){\phantom{$X_1$}0};

        \node[numbox] at (-4,-2.2)  {\phantom{$X_1$}1};
        \node[numbox] at (-4,-2.8)   {\phantom{$X_1$}2};
        \node[numspc] at (-4,-3.4) {$\vdots$};
        \node[numbox] at (-4,-4){\phantom{$X_1$}$U^\beta$};

        \node[numbox] at (-6.5,-2.2)  {$U^\gamma$};
        \node[numbox] at (-6.5,-2.8)  {$U^\gamma$};
        \node[numspc] at (-6.5,-3.4) {$\vdots$};
        \node[numbox] at (-6.5,-4)  {$U^\gamma$};

        \draw[black!75] (-10.5, -7.3) -- (2, -7.3);

        \node[numbox] at (0,-7.8) {$U^\beta$};
        \node[numboxs] at (-2,-7.8) {\phantom{$X_1$}$k$};
        \node[numbox] at (-4,-7.8)  {\phantom{$X_1$}$T + \small\smi{\position{U^\beta}}$};
        \node[numboxl] at (-7.5,-7.8) {$k + U^{\beta + \gamma}$};
    \end{tikzpicture}
    \caption{An overview over the reduction from \kSUM to \SubsetSum. The given \kSUM instance is $(Z,T)$, and we write $Z = \{z_1,\ldots,z_{|Z|}\}$. Bit blocks of the constructed numbers are depicted as boxes, the value of a bit block
    is written inside the corresponding box. The constructed target number is visualized at the
    bottom. The annotations at the top represent the maximum value of the
    constructed numbers up to the specified point. The annotations on the right denote group of constructed numbers. We remark that the number $T + \sm{\position{U^\beta}}$ not necessarily fits in its block, but this is the only overflow that can occur in this figure.}\label{fig:red}
\end{figure}
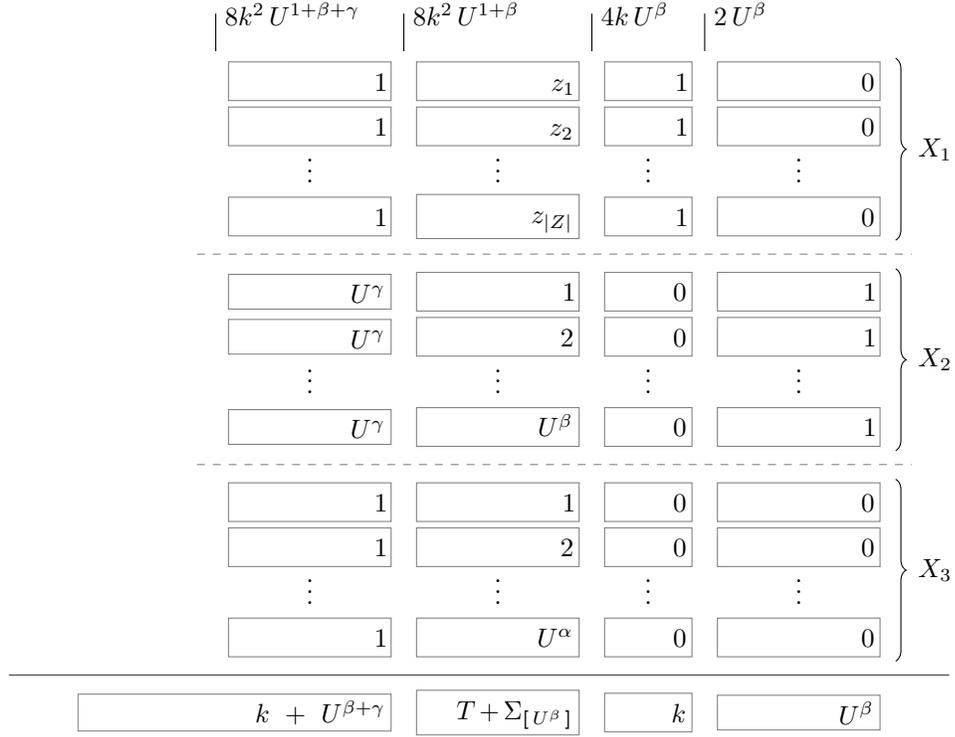
\begin{proof}[Proof of Lemma~\ref{lem:thereduction}]
    We set 
    \[ \beta := \alpha \sigma - \alpha - 1 \qquad \text{and} \qquad \gamma := \alpha(\xi-\sigma+1). \]
    Given a \kSUM instance $(Z, T)$, we construct the following \SubsetSum
    instance $(X,t)$.
    Consider \cref{fig:red} for a visualization.
    \begin{align*}
       X :=& X_1 \cup X_2 \cup X_3, \text{ where} \\
       &X_1 := \{\Const U^{1 + \beta} + z \cdot \const U^\beta + \constt U^\beta\mid z \in Z \}\\
       &X_2 := \{ U^\gamma \cdot \Const U^{1 + \beta} + j \cdot \const U^\beta + 1\mid j \in
            \position{U^\beta}\}\\
       &X_3 := \{ \Const  U^{1 + \beta} + j \cdot \const U^\beta\mid j \in
            \position{U^\alpha}\}\\
        t :=& (k + U^{\beta + \gamma}) \cdot \Const U^{1 + \beta}
        + (T + \smi{\position{U^\beta}}) \cdot \const U^\beta
        + k \cdot \constt U^{\beta}
        + U^\beta
    \end{align*}
    For simplicity, here we assumed that $U^\alpha, U^\beta, U^\gamma$ are integers, more precisely they should be replaced by $\lceil U^\alpha \rceil, \lceil U^\beta \rceil, \lceil U^\gamma \rceil$.
    For this construction to make sense we need $\alpha,\beta,\gamma \ge 0$; we will take care of these bounds later. 
    
    We first verify that the \SubsetSum instance $(X, t)$ is indeed equivalent to
    the \kSUM instance $(Z, T)$.
    \begin{claim} \label{cla:lb_correctness}
        Any solution to the \SubsetSum instance $(X, t)$ corresponds to a solution to
        the \kSUM instance $(Z, T)$, and vice versa.
    \end{claim}
    \begin{claimproof}
        We start with the easier direction: Any solution
        to $(Z, T)$ corresponds to a solution to $(X, t)$.
        To that end, let $B \subseteq Z$ denote a solution to the \kSUM instance
        $(Z, T)$, that is, we have $\smi{B} = T$ and $|B| = k$.
        Consider the set $A \subseteq X$ defined by picking the subset of $X_1$ corresponding to $B$, and picking all numbers in $X_2$, that is,
        \[
            A \;:=\; \{\Const U^{1 + \beta} + z \cdot \const U^{\beta} + \constt U^\beta \mid z \in B\}
            \;\cup\; \{ U^\gamma \cdot \Const U^{1 + \beta} + j \cdot \const U^\beta + 1\mid j \in
            \position{U^\beta}\}.
        \]
        Observe that we have
        \begin{align*}
            \smi{A} &= (k\cdot \Const U^{1+\beta} + T\cdot \const U^{\beta} + k \cdot \constt U^\beta)
            + (U^{\beta + \gamma} \cdot \Const U^{1 + \beta} + \smi{\position{U^\beta}} \cdot \const U^\beta +
            U^\beta)\\
                    &= (k + U^{\beta + \gamma}) \cdot \Const U^{1 + \beta}
                    + (T + \smi{\position{U^\beta}}) \cdot \const U^\beta
                    + k \cdot \constt U^{\beta}
                    + U^\beta
                    = t,
        \end{align*} completing the proof of the first direction.

        For the other direction, let $A \subseteq X$ denote a solution to the \SubsetSum instance $(X, t)$, that is, we have $\smi{A} = t$.
        By construction, we have $t \equiv U^\beta\pmod{\constt U^\beta}$. 
        Since all numbers in $X_1 \cup X_3$ are 0 modulo $\constt U^\beta$, and all $U^\beta$ many numbers in $X_2$ are 1 modulo $\constt U^\beta$, the set $A$ must contain all numbers in $X_2$.
        
        Thus, consider the remaining set $A' := A \setminus X_2$. We have
        \begin{align}
            t' := \smi{A'} = \sm{A} - \sm{X_2} &= t - (U^{\beta + \gamma} \cdot \Const U^{1 + \beta} + \smi{\position{U^\beta}} \cdot \const U^\beta +
            U^\beta) \notag \\
            &= k \cdot \Const U^{1 + \beta}
            + T \cdot \const U^\beta
            + k \cdot \constt U^{\beta}. \label{eq:tprime}
        \end{align}
        Observe that we have $t' \equiv k \cdot \constt U^\beta \pmod{\const U^\beta}$.
        Since the numbers in $X_3$ are 0 modulo $\const U^\beta$, and the numbers in $X_1$ are $\constt U^\beta$ modulo $\const U^\beta$, it follows that $|A' \cap X_1| \equiv k \pmod{2k}$. In particular, we have 
        \begin{align}
          |A' \cap X_1| \ge k. \label{eq:eogsg}
        \end{align}
        
        We can assume without loss of generality that $T \le kU$. This implies 
        \[ t' = k \cdot \Const U^{1 + \beta}
            + T \cdot \const U^\beta
            + k \cdot \constt U^{\beta} < (k+1) \Const U^{1+\beta}. \]
        Since all numbers in $X_1 \cup X_3$ are bounded from below by $\Const U^{1+\beta}$, the bound on $t'$ implies that we can choose at most $k$ items from $X_1 \cup X_3$, that is, $|A'| \le k$. 
        Together with inequality~(\ref{eq:eogsg}), it follows that $A' \subseteq X_1$ and $|A'| = k$. 
        
        So let $B \subseteq Z$ be the subset corresponding to $A' \subseteq X_1$. Then we have
        \[ \sm{A'} = k \cdot \Const U^{1+\beta} + \sm{B} \cdot \const U^\beta + k \cdot \constt U^\beta. \]
        Comparing with (\ref{eq:tprime}), we obtain $\sm{B} = T$.
        Hence, if the \SubsetSum instance $(X,t)$ has a solution $A$, then the \kSUM instance $(Z,T)$ has a solution $B$. This completes the proof of the second
        direction and thus the proof of the claim.
    \end{claimproof}

    We next verify that $\alpha,\beta,\gamma \ge 0$, in addition to other inequalities that we will need in the following.
    \begin{claim} \label{cla:ineq}
      The parameters $\alpha,\beta,\gamma$ satisfy the following inequalities.
    \begin{enumerate}
      \item $0 < \alpha < 1$,
      \item $\beta \ge 0$,
      \item $\gamma \ge 0$,
      \item $\beta + \gamma \le \alpha$,
      \item $\beta \le \alpha$.
    \end{enumerate}
    \end{claim}
    \begin{claimproof}
      (1.) Follows from the parameter setting $(\tau,\xi,\sigma)$ being hard.
      (2.) The non-triviality assumption $\tau \ge \xi$ yields $\sigma - 1 \ge \sigma + \xi - \tau - 1 = 1/\alpha$. After rearranging, we obtain $0 \le \alpha \sigma - \alpha - 1 = \beta$.
      (3.) The non-triviality assumption $\sigma \le 1 + \xi$ yields $\xi - \sigma + 1 \ge 0$. Together with $\alpha > 0$ we obtain $0 \le \alpha (\xi - \sigma + 1) = \gamma$.
      (4.) $\beta + \gamma \le \alpha$: The non-triviality assumption $\tau \le \sigma$ yields $\xi-1 \le \sigma + \xi - \tau - 1 = 1/\alpha$. Rearranging this, we obtain $\alpha \ge \alpha \xi - 1 = \beta + \gamma$.
      (5.) Follows from the preceeding inequality and $\gamma \ge 0$. 
    \end{claimproof}

    It remains to verify that the instance $(X, t)$ belongs to the parameter setting \SubsetSumN$(\tau,\xi,\sigma)$ and fulfills the claimed size bound $|X| = \Theta(U^\alpha)$. 
    \begin{claim} \label{cla:lb_sizes}
        The {\sf Subset Sum} instance $(X, t)$ satisfies
        $n := |X| = \Theta(U^{\alpha})$,
            $\mxi{X} = \Theta(n^\sigma)$,
            $\smi{X} = \Theta(n^{\xi})$, and
            $t = \Theta(n^\tau)$.
    \end{claim}
    \begin{claimproof}
        For the size of $X$ we bound $U^\alpha \le |X| \le U^\alpha + U^\alpha + U^\beta \le
            3 U^\alpha$, 
        where we used $\beta \le \alpha$ (Claim~\ref{cla:ineq}.5).
        
        Note that in the definition of $X_1,X_2,X_3$ the leftmost summand is always asymptotically dominating. In particular, every $x \in X_1 \cup X_3$ satisfies $x = \Theta(U^{1+\beta})$ and every $x \in X_2$ satisfies $x = \Theta(U^{1+\beta+\gamma})$. (Here we treat $k$ as a constant, and for $X_1$ we use  $Z \subseteq \position{U}$, for $X_2$ we use $\beta \le \alpha \le 1$, and for $X_3$ we use $\alpha \le 1$.)
        
        This allows us to determine the maximum number as $\mx{X} = \Theta(U^{1+\beta+\gamma})$. From the definition of $\beta,\gamma$ we see that $1+\beta+\gamma = \alpha \xi$. Hence, $\mx{X} = \Theta(U^{\alpha \xi}) = \Theta(n^\xi)$.
        
        Note that $|X_1| = |Z| \le U^\alpha$, $|X_2| = U^\beta$, and $|X_3| = U^\alpha$. From the resulting $|X_1 \cup X_3| = \Theta(U^\alpha)$ and our bounds on numbers in $X_1 \cup X_3$ and $X_2$, we determine the sum of all numbers in $X$ as
        \[ \sm{X} = \Theta(U^\alpha \cdot U^{1+\beta} + U^\beta \cdot U^{1+\beta+\gamma}). \]
        The inequality $\beta+\gamma \le \alpha$ (Claim~\ref{cla:ineq}.4) now yields $\sm{X} = \Theta(U^{1+\alpha+\beta})$. From the definition of $\beta$, we see that $\sm{X} = \Theta(U^{\alpha \sigma}) = \Theta(n^\sigma)$. 
        
        Finally, we turn to the target $t$. 
        We claim that $t = \Theta(U^{\beta + \gamma} \cdot U^{1+\beta})$, that is, $t$ is asymptotically dominated by its first summand. This is clear for almost all summands. For the summand $T \cdot \const U^\beta$ we use $T \le kU = \Oh(U)$ to see that it is $\Oh(U^{1+\beta})$ and thus dominated by the first summand. For the summand $\sm{\position{U^\beta}} \cdot \const U^\beta$ we use $\sm{\position{U^\beta}} = \Oh(U^{2\beta})$ to see that it is $\Oh(U^{3 \beta}) = \Oh(U^{1+2\beta})$, since $\beta \le \alpha \le 1$, so this summand is also dominated by the first one.
        
         It remains to analyze the exponent of $t = \Theta(U^{1+2\beta+\gamma})$. First plugging in the definitions of $\beta$ and $\gamma$, and then expanding $1 = \alpha/\alpha = \alpha(\xi+\sigma-\tau-1)$, we obtain
         \[ 1+2\beta+\gamma = \alpha (\sigma + \xi - 1) - 1 = \alpha (\sigma + \xi - 1) - \alpha(\xi+\sigma-\tau-1) = \alpha \tau. \]
         Hence, we have $t = \Theta(U^{1+2\beta+\gamma}) = \Theta(U^{\alpha \tau}) = \Theta(n^\tau)$, completing the proof of the claim.
    \end{claimproof}

    From the size bound $|X| = \Theta(U^\alpha)$ and the easy structure of $X$ and $t$, it follows that $(X,t)$ can be computed in time $\Oh(U^\alpha)$.
    Together with Claims~\ref{cla:lb_sizes} and \ref{cla:lb_correctness}, this finishes the proof of Lemma~\ref{lem:thereduction}.
\end{proof}

\section{Conclusion and Open Problems} \label{sec:conclusion}

In this paper we designed improved algorithms and lower bounds for dense \SubsetSum with respect to the parameters $n,t,\mx{X},\sm{X}$. When the input $X$ is a set, we showed a dichotomy into parameter settings where \SubsetSum can be solved in near-linear time $\tOh(n)$ and settings where it cannot, under standard assumptions from fine-grained complexity theory. We also generalized our algorithms to multi-sets. We conclude with some open problems.

\medskip
In the set case, our lower bound characterizes all near-linear time settings, but it does not match the known upper bounds in the super-linear regime. It would be plausible that \SubsetSum can be solved in time $\tOh(n + \min\{t, \mx{X}\sm{X}/(nt)\})$, which would match our lower bound. So far, this running time can be achieved for $t = \tOh(\sqrt{\mx{X} \sm{X}/n})$~\cite{Bringmann17} or $t \gg \mx{X} \sm{X} / n^2$ (Theorem~\ref{thm:intro_main_algo}).

However, this is a hard open problem, since a matching algorithm (or a higher lower bound) would also answer the open problem from~\cite{AxiotisBJTW19} whether \SubsetSum can be solved in time $\tOh(n + \mx{X})$. Indeed, bounding $\sm{X} \le n \cdot \mx{X}$ and $\min\{t, \mx{X}^2 / t\} \le \mx{X}$ we obtain time $\tOh(n + \min\{t, \mx{X}\sm{X}/(nt)\}) = \tOh(n + \mx{X})$.

\medskip
We generalized our algorithm to the multi-set case, at the cost of a factor $\mul{X}$ in the feasibility bound. Generalizing our lower bounds and gaining a similar factor $\mul{X}$ seems complicated. We therefore leave it as an open problem to determine the near-linear time regime in the case of multi-sets.

\medskip
Galil and Margalit's algorithm can be phrased as a data structure: We can preprocess $X$ in time $\tOh(n + \mx{X}^2/n^2)$ so that given a target $t \gg  \mx{X} \sm{X} / n^2$ we can decide whether some subset of $X$ sums to $t$ in time $\Oh(1)$. Our algorithm from Theorem~\ref{thm:intro_main_algo} can be phrased in the same data structure setting, with an improved preprocessing time of $\tOh(n)$. 

However, Galil and Margalit's data structure can even reconstruct solutions, namely after preprocessing~$X$ and given $t$ they can compute a subset $Y \subseteq X$ summing to $t$, if it exists, in time $\tOh(|Y|)$. We leave it as an open problem to extend our algorithm to admit this type of solution reconstruction, as we focused on the decision problem throughout this paper.

\bibliographystyle{abbrv}
\bibliography{ms}

\end{document}